\documentclass[12pt]{article}
\usepackage{amsmath,amssymb,exscale,color,multirow,graphicx,epsfig,mathtools,cite}
\input epsf
\newcommand{\pslash}{{\not \!p}}
\newcommand{\Dslash}{{\not \!\!D}}

\begin{document}
\topmargin -1.0cm
\oddsidemargin -0.8cm
\evensidemargin -0.8cm

\vspace{40pt}

\begin{center}
\vspace{40pt}

\Large \textbf{A vector leptoquark for the $B$-physics anomalies \\ from a composite GUT}

\end{center}

\vspace{15pt}
\begin{center}
{\bf Leandro Da Rold$^{\circ}$, Federico Lamagna$^{\star}$} 

\vspace{20pt}

\textit{Centro At\'omico Bariloche, Instituto Balseiro and CONICET}
\\[0.2cm]
\textit{Av.\ Bustillo 9500, 8400, S.\ C.\ de Bariloche, Argentina}

\end{center}

\vspace{20pt}
\begin{center}
\textbf{Abstract}
\end{center}
\vspace{5pt} {\small \noindent
A vector leptoquark at the TeV scale, mostly coupled to the fermions of the third generation, is the preferred option to explain the hints of lepton flavor universality violation in the decays of $B$-mesons. 
It seems interesting to assume that this leptoquark belongs to the same beyond the Standard Model sector that solves the hierarchy problem, since the third generation of fermions play the leading role in the instability of the Higgs potential.
We present a composite Grand Unified Theory with resonances at the TeV that contains the required vector leptoquark and develops the Higgs as a pseudo Nambu-Goldstone boson. 
We show that anarchic partial compositeness of the Standard Model fermions can accommodate the couplings of Left-handed currents required by the $B$-anomalies, predicting very small couplings to the Right-handed currents without any additional hypothesis.
By making use of an effective theory description of the strong dynamics, in terms of weakly coupled resonances, we are able to compute the corrections to $B$-physics, as well as the one-loop potential for the pseudo Nambu-Goldstone bosons.
The theory has a rich phenomenology and a candidate for dark matter.
}

\vfill
\noindent {\footnotesize E-mail:
$\circ$ daroldl@cab.cnea.gov.ar,
$\star$ federico.lamagna@cab.cnea.gov.ar
}

\noindent
\eject

\tableofcontents

%%%%%%%%%%%%%%%%%%%%%%%%%%%%%%%%%%%%%%%%%%%%%%%%%%%%%%%%%%%%%%%%
\begin{flushright}
{\it We dedicate this work to the memory of Eduardo Pont\'on, \\ great physicist and greater person, who left us too soon.}
\end{flushright}

\section{Introduction}
In the last years different experiments have reported hints of violation of lepton flavor universality (LFU) in semileptonic decays of $B$-mesons, both in the charged current process $b\to c\ell\nu$~\cite{Lees:2013uzd,Aaij:2015yra,Hirose:2016wfn,Aaij:2017deq,Abdesselam:2019dgh}, and in the neutral current process $b\to s\ell\bar\ell$~\cite{Aaij:2014ora,Aaij:2017vbb,Aaij:2019wad,Abdesselam:2019wac}. Although there is no conclusive evidence of new physics yet, the deviations from the predictions of the Standard Model (SM) are above $3\sigma$ for each kind of interaction, leading to one of the most interesting challenges of flavor physics. There is no evidence of a common origin for both deviations, but at theoretical level it is very interesting to explore this possibility. These anomalies could be explained by introducing new physics at a few TeV scale, with interactions having a non-trivial flavor structure, mainly coupled to the fermions of the third generation. 

Many references have shown that the deviations in $B$-physics can be explained by adding to the SM a spin one leptoquark, known in the literature as $U_1$~\cite{Dorsner:2016wpm}, transforming as $({\bf 3},{\bf1})_{2/3}$ under the SM gauge group, with a mass of order few TeV and interactions with Left-handed currents of SM fermions~\cite{Alonso:2015sja,Calibbi:2015kma,Fajfer:2015ycq,Barbieri:2015yvd,Hiller:2016kry,Bhattacharya:2016mcc,Buttazzo:2017ixm,Assad:2017iib,
DiLuzio:2017vat,Calibbi:2017qbu,Bordone:2017bld,Barbieri:2017tuq,Blanke:2018sro,Greljo:2018tuh,Bordone:2018nbg,Matsuzaki:2018jui,
Kumar:2018kmr,Crivellin:2018yvo,DiLuzio:2018zxy,Angelescu:2018tyl,Balaji:2018zna,Fornal:2018dqn,Ciuchini:2019usw,Cornella:2019hct,Crivellin:2019szf}. Refs.~\cite{Buttazzo:2017ixm,Crivellin:2018yvo,Angelescu:2018tyl} have made a detailed analysis of flavor observables, showing what kind of couplings can explain the anomalies and simultaneously satisfy the bounds from other flavor observables. This scenario remains as the best one to explain the $B$-physics puzzle with a single new particle at the TeV scale.~\footnote{See also, for example, Ref.~\cite{Popov:2019tyc} for a solution with a single scalar leptoquark $R_2$.}

As is well known new dynamics at the TeV scale, mainly coupled to the third generation, is also needed to stabilize the Higgs potential. Although there is no obvious connection between these deviations and the hierarchy problem, from a theoretical perspective, it would be very interesting to find a common origin for both phenomena.

Grand Unified Theories (GUT) naturally predict the presence of leptoquarks, but usually with masses at the scale of grand unification. However, in composite GUTs, where a new strongly coupled field theory (SCFT) is introduced, the compositeness scale can be taken at the TeV scale, leading to leptoquark resonances with masses of few TeV. In composite GUTs usually the grand unified group H is a global symmetry of the SCFT, containing as a subgroup the gauge symmetry of the SM (G$_{\rm SM}$). In these scenarios the gauge and fermion fields of the SM are taken as elementary fields, weakly coupled to the SCFT, and gauging the subgroup G$_{\rm SM}$ of H. Composite GUTs can also solve the hierarchy problem: if the SCFT has a larger group G, spontaneously broken to H, the Higgs can emerge as a composite Nambu-Goldstone Boson (NGB) state in G/H, as discussed for example in Ref.~\cite{Frigerio:2011zg}.~\footnote{In this kind of composite GUTs, the global subgroup H must also contain the custodial symmetry of the Higgs sector to have a chance to pass the electroweak precision tests (EWPT).} As usual in composite Higgs models, since the elementary fields do not furnish full representations of the global symmetry of the SCFT, at loop level they induce a potential for the NGBs. Under some suitable conditions this potential can trigger electroweak symmetry breaking (EWSB) dynamically.

In the present work we do not pursue precise gauge coupling unification. Instead we are guided by the low energy phenomenology, and we demand G to be such that the SCFT contains spin one resonances with the proper quantum numbers to be identified with $U_1$, as well as a NGB Higgs. A solution to the $B$-anomalies, besides a $U_1$, requires a well defined flavor structure of couplings. One of the most interesting approaches to flavor physics in composite models is anarchic partial compositeness, where the interactions of the elementary fermions with the SCFT are dominated by linear interactions. As is well known, in this case a hierarchy of elementary-composite linear mixings can be generated at low energies by the running of the linear couplings, leading to the hierarchical spectrum and mixing angles of the CKM matrix, and simultaneously suppressing the flavor violating processes.
Since the same elementary-composite mixings enter in the interactions with all the resonances, once the Left-handed mixings of the leptons are fixed to explain the $B$-anomalies, the Right-handed ones are fixed to obtain the Yukawa couplings. As $R_{D^{(*)}}$ requires a large mixing for $\tau_L$, the resulting Right-handed mixings are suppressed by the ratio of charged lepton mass over the Higgs vacuum expectation value (vev), giving very small Right handed couplings with $U_1$. Thus anarchic partial compositeness gives, as a very good approximation, interactions of $U_1$ with Left-handed currents and negligible interactions with Right handed currents, without any additional hypothesis.

An effective weakly coupled description of the above dynamics can be obtained by working with a theory of resonances. We will consider a three-site theory, with the first site describing the elementary sector and the other two sites describing resonances of the SCFT. In this case the one-loop potential of the NGBs is finite and can be calculated explicitly, as well as the spectrum of new states and their couplings, leading to well defined predictions. We will show that composite GUTs can simultaneously explain the $B$-anomalies and stabilize the Higgs potential. Besides, due to the large degree of compositenes required for $\tau_L$, the third generation of Left-handed leptons play an important role in the potential. This situation was considered in Ref.~\cite{Carmona:2013cq}, although in a different context.

Our paper is organized as follows: in sec.~\ref{sec-cGUT} we show a composite GUT containing the usual ingredients of composite Higgs models, as well as a vector leptoquark for the $B$-anomalies. We describe the coset structure of the SCFT, the content of NGBs, the fermionic representations and flavor structure, as well as some important bounds and estimates associated to $B$-physics, as $R_{D^{(*)}}$ and $R_{K^{(*)}}$. In the same section we present the effective low energy physics obtained after integration of the massive resonances of the SCFT, whose structure depends only on the pattern of symmetries, and the one-loop potential of the NGBs. In sec.~\ref{sec-3sites} we present an effective description of the resonances of the SCFT in terms of a three-site model. In sec.~\ref{sec-pheno} we describe the phenomenology of the theory, we scan the parameter space finding regions with EWSB and we compute the spectrum of new particles. We also calculate the corrections to several observables, comparing them with the present bounds, as well as the corrections to flavor quantities as $R_{D^{(*)}}$. Finally, we comment very briefly on the phenomenology of the new pseudo Nambu Goldstone boson (pNGB) states. We present our conclusions and some discussions in sec.~\ref{sec-conclusions}. In an appendix we present a 5D model that can also be used to describe the SCFT.

%%%%%%%%%%%%%%%%%%%%%%%%%%%%%%%%%%%%%%%%%%%%%%%%%%%%%%%%%%%%%%%%
\section{A Composite GUT for the $B$-anomalies}\label{sec-cGUT}
We consider a theory with two different sectors: an SCFT or composite sector and another sector called elementary that is weakly coupled with the SCFT. 
The SCFT is assumed to have a simple global symmetry group G, spontaneously broken by the strong dynamics to a subgroup H. This breaking generates a set of NGBs associated to the broken generators of the coset G/H. Some of these NGBs will be identified as a composite Higgs. The conserved Noether currents of the SCFT can create resonances of spin one, transforming with the adjoint representation of G. Besides, we also assume that the there are fermionic operators ${\cal O}^{\rm SCFT}$ that can create spin 1/2 resonances, transforming with linear irreducible representations of G. These representations are not fixed {\it a priori}, leading to some freedom for model building. The masses of the first level of resonances, collectively denoted as $m_*$, are taken of order few TeV, whereas the interactions between them are characterized by a single coupling $g_*$, taken as: $g_{\rm SM}\ll g_*\ll 4\pi$, thus for simplicity we assume that all the couplings between resonances are of the same order. The NGB decay constant is $f=m_*/g_*$, of order TeV.

The gauge fields and fermions of the SM are external to the SCFT, they are taken as elementary fields. Demanding G to contain the SM gauge symmetry group, the gauging of G$_{\rm SM}$ explicitly breaks the global symmetry of the SCFT. The fermions of the SM have linear interactions with the SCFT, that also break G explicitly~\footnote{We assume that bilinear interactions are suppressed, having no impact in the phenomenology, except possibly for the neutrino sector~\cite{Agashe:2008fe} and eventually the first generation~\cite{DaRold:2017xdm}.}:
\begin{equation}\label{eq-linear}
{\cal L}\supset \omega_\psi\bar\psi{\cal O}^{\rm SCFT}_\psi+{\rm h.c.}\ , 
\end{equation}
with $\omega_\psi$ being the coupling at the high ultraviolet (UV) scale $\Lambda$ at which this Lagrangian is defined.

\subsection{Coset structure}
SO(11)/SO(10) is the minimal coset of simple groups with the following properties: it contains the SM gauge symmetry group as well as custodial symmetry, it delivers a Higgs as a pNGB and, after proper identification of hypercharge, it contains a composite spin one state that has the proper quantum numbers to be identified with the $U_1$ leptoquark (we follow the notation of Ref.~\cite{Dorsner:2016wpm} for leptoquarks).~\footnote{The first two properties were already shown in Ref.~\cite{Frigerio:2011zg}, the last one, as far as we know, has not been considered before.} However, since in this case $U_1$ is associated to a broken generator, it is heavier than, for example, $W'$ and $Z'$ resonances, resulting in a suppressed effect in $R_{K^{(*)}}$ and $R_{D^{(*)}}$, that are proportional to $m_{U_1}^{-2}$, and thus can not be accommodated. For this reason, we will consider instead a larger coset: SO(12)/SO(11), such that $U_1$ can be associated with an unbroken generator. Let us discuss the coset structure in some detail.

We start by describing some features of the unbroken group. SO(11) contains SO(10) that, as is well known from the study of GUTs, can accommodate a Left-Right symmetric extension of the SM gauge group. A possible pattern of subgroups that allows to see this property is: 
\begin{equation}\label{eq-patts}
{\rm SO}(11)\to{\rm SO}(10)\to{\rm SO}(6)\times{\rm SO}(4)\to{\rm SU}(3)_c\times{\rm SU}(2)_L\times{\rm SU}(2)_R\times{\rm U}(1)_X\equiv{\rm H}_{\rm min} \ ,
\end{equation}
with SO(6)$\sim$SU(4)$\supset SU(3)\times$U(1) and SO(4)$\sim$SU(2)$\times$SU(2). Besides, we identify hypercharge with the following combination:
\begin{equation}\label{eq-Y}
Y\equiv T^{3R}+\frac{4}{\sqrt{6}}T^X \ .
\end{equation}

The set of broken generators in the coset SO(12)/SO(11) transform, under SO(11), with the representation {\bf11}. Under SO(10) and H$_{\rm min}$ the representation {\bf11} decomposes as: 
%\begin{equation}
%{\bf10}\sim({\bf3},{\bf1},{\bf1})_{1/\sqrt{6}}\oplus(\bar{\bf3},{\bf1},{\bf1})_{-1/\sqrt{6}}\oplus({\bf1},{\bf2},{\bf2})_0 \ . 
%\end{equation}
\begin{equation}
{\bf11}\sim\bf1\oplus\bf10\sim({\bf1},{\bf1},{\bf1})_0\oplus({\bf1},{\bf2},{\bf2})_0\oplus(\bar{\bf3},{\bf1},{\bf1})_{-1/\sqrt{6}}\oplus{\rm c.c.} \ , 
\label{eq-repNGB}
\end{equation}
where $\oplus$~c.c. means that, for the complex representations as the color triplet, one has to add the charge conjugate one. Eq.~(\ref{eq-repNGB}) shows the transformation properties of the NGBs, those associated to the colorless generators lead to two multiplets: a SM singlet that we call $\varphi$ and the Higgs field $H$, whereas the ones associated to the color triplet lead to a leptoquark usually called $\bar S_1$ in the literature.

The currents of the SCFT associated to the global symmetry SO(12) can create spin one states that transform with the adjoint representation {\bf66}, that under SO(11) decomposes as: $\bf66\sim\bf55\oplus\bf11$. We have shown in Eq.~(\ref{eq-repNGB}) the decomposition of {\bf11}, the representation $\bf55$ decomposes under SO(10) and H$_{\rm min}$ as:
%\begin{equation}
%{\bf45}\sim({\bf3},{\bf1},{\bf1})_{-2/\sqrt{6}} \oplus (\bar{\bf3},{\bf1},{\bf1})_{2/\sqrt{6}} \oplus ({\bf3},{\bf2},{\bf2})_{1/\sqrt{6}} \oplus (\bar{\bf3},{\bf2},{\bf2})_{-1/\sqrt{6}} \oplus ({\bf8},{\bf1},{\bf1})_0 \oplus ({\bf1},{\bf3},{\bf1})_0 \oplus ({\bf1},{\bf1},{\bf3})_0 \oplus ({\bf1},{\bf1},{\bf1})_0 \ . 
%\end{equation}
%
\begin{align}
{\bf55}\sim&{\bf45}\oplus{\bf10}
\nonumber\\
\sim&({\bf8},{\bf1},{\bf1})_0 \oplus ({\bf1},{\bf3},{\bf1})_0 \oplus ({\bf1},{\bf1},{\bf3})_0 \oplus ({\bf1},{\bf1},{\bf1})_0 \oplus ({\bf3},{\bf1},{\bf1})_{-2/\sqrt{6}} \oplus ({\bf3},{\bf2},{\bf2})_{1/\sqrt{6}} 
\nonumber\\
&\oplus({\bf1},{\bf2},{\bf2})_0\oplus({\bf3},{\bf1},{\bf1})_{1/\sqrt{6}}\oplus{\rm c.c.}\ . 
\label{eq-rep55}
\end{align}
where the first line contains the decomposition of the {\bf45}, and the second one of the {\bf10}.
With the identification of hypercharge of Eq.~(\ref{eq-Y}), the multiplets $({\bf3},{\bf1},{\bf1})_{1/\sqrt{6}}\oplus{\rm c.c.}$ contained in the {\bf10} of {\bf55} and {\bf11} can be identified with $U_1$ leptoquarks. The leptoquark in {\bf55} is associated to an unbroken generator, whereas the leptoquark in {\bf11} is associated to a broken one, thus the former results lighter than the latter. 

Besides $U_1$, there is another spin one leptoquark: $\tilde V_2\sim({\bf3},{\bf2})_{1/6}$, as well as two new states transforming as: $({\bf3},{\bf2})_{7/6}$ and $({\bf3},{\bf1})_{-4/3}$. In generic leptoquark models $\tilde V_2$ can induce baryon decay, however, as we will show in sec.~\ref{sec-baryon}, the present model has a global U(1)$_B$ that forbids proton decay. The other two states do not have dimension-four operators with SM fermions.

It is also possible to choose other identifications of hypercharge, as $Y\equiv T^{3R}-2T^X/\sqrt{6}$, that allow to embed $U_1$ in $({\bf3},{\bf1},{\bf1})_{-2/\sqrt{6}}\oplus{\rm c.c.}$. However in this case the NGB leptoquark is an $S_1$, giving contributions to $B$-physics that can destabilize the $U_1$ solution.

We will add a discrete $Z_2$-symmetry, that corresponds to a parity and enlarges SO(12) to O(12). We are interested in the transformation under which broken and unbroken generators are, respectively, odd and even under this parity, leading to odd NGBs. In the basis defined in Ap.~\ref{ap-SO(12)}, for the representation {\bf12} this parity can be written in terms of a $12\times 12$ matrix as: $P_{ij}=\delta_{ij}-2\delta_{i12}\delta_{j12}$. As we will show, the presence of $P$ will lead to several simplifications as well as a candidate for dark matter. 

\subsection{Fermions}\label{sec-ferm}
The operators ${\cal O}^{\rm SCFT}$ that interact linearly with the SM fermions can be decomposed under $G_{\rm SM}$ as sums of irreducible representations. To avoid explicit breaking of G$_{\rm SM}$, these decompositions must contain the representations of the SM fermions. Given Eq.~(\ref{eq-Y}), partners of the SM fermions can be found in the following representations of SO(10):
\begin{equation}
u, \ell \subset {\bf10}\ , \qquad q, e \subset {\bf45} \ ,\qquad q,d,\ell \subset {\bf120} \ .
\label{eq-embed1}
\end{equation}
A Right-handed neutrino can be embedded in a singlet or in the adjoint representation of SO(10). Larger representations are also possible. The representations of Eq.~(\ref{eq-embed1}) can be embedded in representations of SO(12), we are interested in the following:
\begin{align}
%&\bf12 \sim \bf11\oplus\bf1 \sim \bf10\oplus\bf1\oplus\bf1 \ , 
%\nonumber\\
&\bf66 \sim\bf55\oplus\bf11 \sim(\bf45\oplus \bf10')\oplus (\bf10\oplus\bf1) \ , 
\nonumber\\
&\bf220 \sim \bf165\oplus\bf55 \sim (\bf120\oplus\bf45')\oplus(\bf45\oplus \bf10')
\label{eq-embed2}
\end{align}
where we have shown the decompositions under SO(11) and SO(10). We have used the marks to distinguish SO(10) representations that arise from the decomposition of different representations of SO(11).

In order to obtain interactions between all the SM fermions and the SCFT, we will consider that the following operators are present: ${\cal O}^{\rm SCFT}_{\bf66}$ and ${\cal O}^{\rm SCFT}_{\bf220}$.
% and ${\cal O}^{\rm SCFT}_{\bf165}$. Strictly speaking ${\cal O}^{\rm SCFT}_{\bf11}$ is not needed, since ${\cal O}^{\rm SCFT}_{\bf55}$ already contains partners of $u$ and $\ell$, but, as we will discuss in sec.~\ref{sec-ewsb}, it will be useful to add it.
Each elementary fermion can interact with more than one SCFT operator, for example $q$ can interact with ${\cal O}^{\rm SCFT}_{\bf66}$ and ${\cal O}^{\rm SCFT}_{\bf220}$, however we will assume that the SCFT operators have different anomalous dimensions, such that one of the interactions dominates over the other (see sec.~\ref{sec-pcf}), and as a simplification of this situation we will consider that each elementary fermion interacts just with one ${\cal O}^{\rm SCFT}$. In particular, as shown in table~\ref{t-irreps}, we assume that $q$, $u$ and $\ell$ interact with ${\cal O}^{\rm SCFT}_{\bf66}$ only, whereas $d$ and $e$ interact with ${\cal O}^{\rm SCFT}_{\bf220}$.
Besides, from Eqs.~(\ref{eq-embed1}) and (\ref{eq-embed2}) one can see that the elementary fermions $u$ and $\ell$ can interact with several components of ${\cal O}_{\bf 66}^{\rm SCFT}$: either with the ${\bf10}\subset{\bf11}$ or with the ${\bf10}'\subset{\bf55}$.
%, whereas $d$ can interact just with the ${\bf120}\subset{\bf165}$ of ${\cal O}^{\rm SCFT}_{\bf220}$. 
The parity $P$ can distinguish between both {\bf10}s inside {\bf66}: ${\bf10}'$ is even and {\bf10} is odd, thus if we assign a well defined parity to the elementary fermions, $P$ is conserved and the elementary fermions $u$ and $\ell$ interact only with one multiplet of SO(11) in ${\cal O}^{\rm SCFT}_{\bf66}$. In the following we will assign the parities of table~\ref{t-irreps} to the elementary fermions, and we will mix them with the components of the SCFT operators shown in that table. 

\begin{table}[ht]
\centering
\begin{tabular}{|c|c|c|c|c|c|}
\hline\rule{0mm}{5mm}
field & $P$ & $H_{\rm min}$ & SO(10) & SO(11) & SO(12) \\[5pt]
\hline \rule{0mm}{4mm}
$q$ & + & $({\bf3},{\bf2},{\bf2})_{1/\sqrt{6}}$ & {\bf45} & {\bf55} & {\bf66} \\[3pt]
\hline  \rule{0mm}{4mm}
$\ell$ & + & $({\bf1},{\bf2},{\bf2})_{0}$ & {\bf45} & {\bf55} & {\bf66} \\[3pt]
\hline  \rule{0mm}{4mm}
$u$ & - & $({\bf3},{\bf1},{\bf1})_{1/\sqrt{6}}$ & {\bf10} & {\bf11} & {\bf66} \\[3pt]
\hline  \rule{0mm}{4mm}
$d$ & - & $({\bf3},{\bf1},{\bf3})_{1/\sqrt{6}}$ & {\bf120} & {\bf165} & {\bf220} \\[3pt]
\hline  \rule{0mm}{4mm}
$e$ & - & $({\bf1},{\bf1},{\bf3})_{0}$ & ${\bf45}'$ & {\bf165} & {\bf220} \\[3pt]
\hline  \rule{0mm}{4mm}
$H$ & - & $({\bf1},{\bf2},{\bf2})_{0}$ & {\bf10} & {\bf11} & $\times$ \\[3pt]
\hline \rule{0mm}{4mm}
$\bar S_1$ & - & $(\bar{\bf3},{\bf1},{\bf1})_{-1/\sqrt{6}}$ & {\bf10} & {\bf11} & $\times$ \\[3pt]
\hline  \rule{0mm}{4mm}
$\varphi$ & - & $({\bf1},{\bf1},{\bf1})_{0}$ & {\bf1} & {\bf11} & $\times$ \\[3pt]
\hline 
\end{tabular}
\caption{Embedding of the composite partners of the elementary fermions, from $H_{\rm min}$ up to SO(12). In the last three lines of the table we show the NGBs that transform with the fundamental representation of SO(11).}
\label{t-irreps}
\end{table}

It is also interesting to consider the scenario without $P$, we will briefly comment on the consequences of this assumption in the section~\ref{sec-noP}.

Let us now describe the interactions between the elementary fermions and the Higgs. Since bilinear interactions with the Higgs have been assumed to be suppressed, the interactions with the Higgs are mediated by the linear interactions of Eq.~(\ref{eq-linear}). The resonances of the SCFT interact with the composite Higgs and Eq.~(\ref{eq-linear}) leads to Yukawa interactions of the elementary fermions. The SCFT has a global unbroken symmetry SO(11), thus in order to obtain the proper Yukawa interactions of the SM, the interactions between the resonances containing the partners of the SM fermions and the Higgs must be SO(11)-invariant. For the up-type quarks, from the embeddings of table~\ref{t-irreps}: ${\bf55}\times{\bf11}\sim{\bf11}\oplus{\bf165}\oplus{\bf429}$, whereas for the down-type quarks and the charged leptons: ${\bf55}\times{\bf165}\sim{\bf11}\oplus\dots$, thus our choice is compatible with the Higgs embedded in an {\bf11}, and $P$-symmetry is respected by Yukawa interactions. 

Usually the SM fermions are embedded in the representation {\bf32} of SO(11), however in order to do that one has to take a different identification of hypercharge~\cite{Frigerio:2011zg}. It is also possible to take other representations, as {\bf12}, that contains $\ell$ and $u$, both $P$-odd, but for simplicity we will not consider them.

\subsection{Partial compositeness and flavor structure}\label{sec-pcf}
At low energies one can consider an effective description of the SCFT in terms of resonances $\Psi$ (we use capital letters for resonances, and small letters for elementary fields). The linear interactions lead to mixing:
\begin{equation}
{\cal L}_{\rm eff}\supset \lambda_\psi f \bar\psi P_\psi\Psi \ ,
\label{eq-linearint}
\end{equation}
where a sum over $\psi=q,u,d,\ell,e$ must be understood. Since $\psi$ is in a representation of G$_{\rm SM}$ and $\Psi$ is in a representation of G, the GUT symmetry, strictly speaking one has to add the projector $P_\psi$, that when acting on $\Psi$ selects the component with the same quantum numbers as $\psi$, for example: $\bar q P_q\Psi\equiv \bar q \Psi_{({\bf 3},{\bf 2})_{1/6}}$.

Assuming that the running of the couplings is driven by the dimension of operator: $\Delta_\psi$, the coupling can be estimated to scale as: $\lambda_\psi\sim(m_*/\Lambda)^{\Delta_\psi-5/2}$. Thus for $\Lambda\gg m_*$, if $\Delta_\psi>5/2$ the coupling $\lambda_\psi$ is suppressed, whereas for $\Delta_\psi\simeq 5/2$ it is not. In this way a hierarchy of mixings can be obtained for different fermions~\cite{Contino:2004vy}.

The mass eigenstates can be obtained after a rotation of angle: $\tan\theta_\psi=\lambda_\psi f/m_\Psi\equiv\epsilon_\psi$~\cite{Contino:2006nn}, realizing partial compositeness. This rotation leads to a chiral massless state that is partially composite, with degree of compositeness $\epsilon_\psi$, as well as a massive resonance with a chiral component that is partially elementary.

The interactions between the elementary fermions and the Higgs require insertions of $\lambda_\psi$, thus the Yukawa couplings can be estimated as: $y_\psi\sim \epsilon_{\psi_L}g_*\epsilon_{\psi_R}$. When considering generations, the mixings and couplings acquire generation indices: $\lambda_{\psi j}$ and $g_{*jk}$.~\cite{Panico:2015jxa} We will assume that the flavor structure of the SCFT is anarchic, this means that in the SFCT there are no preferred directions in flavor space, therefore all the coefficients of the couplings between fermionic resonances are of the same order: $g_*\times {\cal O}(1)$, as well as their masses: $m_*\times {\cal O}(1)$. The hierarchy of SM fermionic masses is driven by hierarchical mixings, light fermions require at least one of the chiral mixings being small, whereas the top mass requires sizable mixing for both chiralities, since: $1\simeq y_t\sim\epsilon_{q3}g_*\epsilon_{u3}$. The mixings of the quarks can be related with the CKM angles and the quark masses, assuming that each chiral SM multiplet interacts predominantly with just one resonance, one gets:~\cite{Agashe:2004cp,Agashe:2008uz}
\begin{align}
&\epsilon_{q1}\sim\lambda_C^3 \epsilon_{q3}\ , 
\qquad 
\epsilon_{u1}\sim \frac{y_u^{\rm SM}}{\lambda_C^3g_*\epsilon_{q3}} \ , 
\qquad
\epsilon_{d1}\sim \frac{y_d^{\rm SM}}{\lambda_C^3g_*\epsilon_{q3}} \ ,
\nonumber \\
&\epsilon_{q2}\sim\lambda_C^2 \epsilon_{q3}\ ,
\qquad
\epsilon_{u2}\sim \frac{y_c^{\rm SM}}{\lambda_C^2g_*\epsilon_{q3}}  \ , 
\qquad
\epsilon_{d2}\sim \frac{y_s^{\rm SM}}{\lambda_C^2g_*\epsilon_{q3}} \ ,
\nonumber \\
&
\qquad\qquad\qquad\qquad
\epsilon_{u3}\sim \frac{y_t^{\rm SM}}{g_*\epsilon_{q3}} \ ,
\qquad\quad
\epsilon_{d3}\sim \frac{y_b^{\rm SM}}{g_*\epsilon_{q3}} \ ,
\label{eq-mixq}
\end{align}
where $y_f^{\rm SM}$ is the SM Yukawa coupling of the quark $f$ and $\lambda_C\simeq 0.22$ is the Cabibbo angle. $\epsilon_{q3}$ and $g_*$ are not fixed by these equations.

The lepton sector depends on the nature of the neutrino and the realization of their masses. The masses of the charged leptons require:
\begin{equation}
\epsilon_{\ell 1}g_*\epsilon_{e1}\sim y_{e}^{\rm SM} \ ,\qquad
\epsilon_{\ell 2}g_*\epsilon_{e2}\sim y_{\mu}^{\rm SM} \ ,\qquad
\epsilon_{\ell 3}g_*\epsilon_{e3}\sim y_{\tau}^{\rm SM} \ ,
\label{eq-mixl}
\end{equation}
As we will show in sec.~\ref{sec-Ba}, in the present scenario the $B$-anomalies can be fitted with a hierarchical mixing of the Left-handed leptons: $\epsilon_{\ell 1}\ll\epsilon_{\ell 2}\ll\epsilon_{\ell 3}$. We will also show in that section that, for the given values of $\epsilon_{\ell i}$, the mixings of the Right-handed charged leptons are also hierarchical: $\epsilon_{e 1}\ll\epsilon_{e 2}\ll\epsilon_{e 3}$, and besides, at least for the second and third generations, they are smaller than the corresponding Left-handed ones: $\epsilon_{e 2}\ll\epsilon_{\ell 2}$ and $\epsilon_{e3}\ll\epsilon_{\ell 3}$. Since these hierarchical mixings lead to small mixing angles in the matrices diagonalizing the charged mass matrix, the large mixing angles of the PMNS matrix must be generated in the neutrino sector. We will assume this to be the case, and we will not elaborate more on the neutrino masses, see Refs.~\cite{Agashe:2008fe,Panico:2015jxa,DaRold:2017xdm} for some examples.

The interactions with the spin one resonances have a flavor structure similar to the Yukawa couplings, except that in this case the factor $g_*$ is universal, due to the global symmetry of the SCFT, thus generically they are misaligned with the Yukawa couplings. The interactions with $U_1$ leptoquarks are of special interest for our analysis:
\begin{equation}
{\cal L}\supset g_{Ljk}^{(n)}\bar q_L^j \gamma^\mu U_{1\mu}^{(n)}\ell_L^k \ + g_{Rjk}^{(n)}\bar d_R^j \gamma^\mu U_{1\mu}^{(n)} e_R^k \ ,%+ \tilde g_{Rjk}\bar u_R^j \gamma_\mu U_1^\mu \nu_R^k \  
\label{eq-LU1}
\end{equation}
where the index $n$ numerates the $U_1$ states, a sum over $n$ is understood. The couplings can be estimated as:
\begin{equation}
g_{Ljk}^{(n)}\sim \frac{c_{jk}}{\sqrt{2}}\epsilon_{qj}g_*\epsilon_{\ell k} \ , \qquad g_{Rjk}^{(n)}\sim \frac{c_{jk}}{\sqrt{2}} \epsilon_{dj}g_*\epsilon_{e k} \ ,
\label{eq-gU1}
\end{equation}
where the factor $1/\sqrt{2}$ arises from the SO(11) generators, and the factor $c_{jk}\sim{\cal O}(1)$.

Using the estimates of Eq.~(\ref{eq-mixq}) for $\epsilon_{dj}$, as well as the ones of sec.~\ref{sec-Ba} for $\epsilon_{ej}$, the couplings $g_{Rjk}^{(n)}$ become very suppressed, and the Right-handed interactions of the second term of Eq.~(\ref{eq-LU1}) can be safely ignored. See Ap.~\ref{ap-coup} for their numerical estimates.

\subsection{Baryon and lepton number conservation}\label{sec-baryon}
Leptoquarks can mediate baryon decay making the theory phenomenologically unacceptable, unless they have very large masses, typically of order $\sim 10^{16}$~GeV.~\footnote{In fact this scale depends on the nature of the leptoquark, as well as on the size of its couplings to the SM fermions.} In the present model there are, for example, vector leptoquarks $\tilde V_2$, with masses of order few TeV, that in principle could couple to diquarks inducing baryon decay. However the SO(11) subgroup contains a generator that can be identified with an operator of baryon number: $B=\sqrt{2/3}T^X$, with $T^X$ the generator of the U(1)$_X$ defined in Eq.~(\ref{eq-patts}). This symmetry assigns the expected baryon number to the resonances, and acts in the usual way on the elementary states, forbidding the  coupling of leptoquarks to diquarks and ensuring baryon number conservation. Thus in the present model $Y=T^{3R}+2B$.

As discussed in Ref.~\cite{Frigerio:2011zg}, the Weinberg dimension five operator can be induced, with a Wilson coefficient that can be generically estimated to be of order $\epsilon_\ell^2/m_*$, resulting in a too large contribution to neutrino masses. To avoid these contributions one can add a U(1)$_L$ global symmetry to the composite sector, assigning the usual numbers to the operators mixing with the elementary fields, for example: $L{\cal O}^{SCFT}_\ell={\cal O}^{SCFT}_\ell$ and $L{\cal O}^{SCFT}_q=0$.

\subsection{$B$-anomalies}\label{sec-Ba}
In order to study the $B$-physics it is convenient to work with the effective theory resulting from the tree-level integration of the resonances. Except where explicitly stated, we will closely follow the analysis of Ref.~\cite{Buttazzo:2017ixm}. As discussed in sec.~\ref{sec-pcf}, only the effect of the $U_1$ leptoquarks on Left-handed currents is important in our model. We obtain the following effective Lagrangian
\begin{equation}\label{eq-l4f}
{\cal L}_{\rm eff}\supset\frac{C^{ijkl}}{v^2_{\rm SM}}[(\bar q^i_L\gamma^\mu\sigma^aq^j_L)(\bar \ell^k_L\gamma_\mu\sigma^a\ell^l_L)+(\bar q_L^i\gamma^\mu q_L^j)(\bar \ell_L^k\gamma_\mu\ell_L^l)] \ ,
\end{equation}
with $i,j,k,l$ being generation indices. The dimensionless coefficient $C^{ijkl}$ is given by:
\begin{equation}\label{eq-C3}
C^{ijkl}=g_{Lil}^{(n)}g_{Ljk}^{(n)*}\frac{v_{\rm SM}^2}{2m_{U_1}^{(n)2}}\sim c_{il}c_{jk}^* \epsilon_{qi}\epsilon_{qj}\epsilon_{\ell k}\epsilon_{\ell l}\frac{v_{\rm SM}^2}{f^2} \ ,
\end{equation}
where we have used that: $m_{U_1}\simeq g_*f/\sqrt{2}$, and we have assumed that the contribution from the lightest
resonance dominates the sum, as we will show that happens in a three-site model. Below we estimate the contributions of our model to $B$-physics, and in sec.~\ref{sec-RDn} we show the numerical predictions in a three-site model.

The SM prediction~\cite{Bigi:2016mdz,Bernlochner:2017jka,Jaiswal:2017rve,Bigi:2017jbd} and the experimental value of $R_{D^{(*)}}^{\tau\ell}\equiv R_{D^{(*)}}$, including the recent results of Belle~\cite{Abdesselam:2019dgh,Abdesselam:2019wac}, are:
\begin{align}
&R_{D}=0.297\pm 0.015 \ ,
\qquad
R_{D^{*}}=0.334\pm 0.031 \ ,\nonumber
\\
&R_{D}^{\rm SM}=0.299\pm 0.003 \ ,
\qquad
R_{D^{*}}^{\rm SM}=0.258\pm 0.005 \ ,
\label{eq-RD0}
\end{align}
 Eq.~(\ref{eq-l4f}) gives a contribution to $R_{D^{(*)}}$ that, to linear order in $C$, can be approximated by:
\begin{equation}\label{eq-RD1}
\frac{R_{D^{(*)}}}{R_{D^{(*)}}^{\rm SM}}\simeq 1+2 C^{3233}\left(1-\frac{V_{tb}^*}{V_{ts}^*}\frac{g_{L23}}{g_{L33}}\right)
\end{equation}
Using the estimates of Eq.~(\ref{eq-mixq}) for the quark degree of compositeness, a fit of $R_{D^{(*)}}$ requires $c_{23}g_{L33}/m_{U_1}\sim 1/$TeV, with $c_{23}\sim {\cal O}(1)$ arising from the dependence of the Wilson coefficient on the coupling $g_{L23}$. In our model this ratio can be estimated as: $\sim\epsilon_{q3}\epsilon_{\ell 3}/f$, therefore we obtain: $\epsilon_{q3}\epsilon_{\ell 3}/f\sim {\cal O}(1)/$TeV. This implies that, for $f\sim$TeV, the Left-handed $\tau$ must have a large degree of compositeness. 

The deviations in $R_{K^{(*)}}$ point to LFU violation in $b\to s\ell\ell$. For negligible coupling to electrons, the preferred contribution from new physics to the Wilson coefficients $\Delta C_9^{\ell\ell}$ and $\Delta C_{10}^{\ell\ell}$ is:~\cite{Alguero:2019ptt,Alok:2019ufo,Aebischer:2019mlg}
\begin{equation}\label{eq-RK0}
\Delta C_9^{\mu\mu}=-\Delta C_{10}^{\mu\mu}=\frac{4\pi}{\alpha_{\rm em}V_{tb}V_{ts}^*}C^{2322}=-0.40\pm 0.12 \ .
\end{equation}
Using Eq.~(\ref{eq-C3}) we obtain: $g_{L32}g^*_{L22}/m_{U_1}^2\simeq 10^{-3}$. Making use of Eq.~(\ref{eq-mixq}), leads to $\epsilon_{q3}\epsilon_{\ell 2}/f\sim 0.1/$TeV, that fixes the order of magnitude of $\epsilon_{\ell 2}$.

As long as $\epsilon_{\ell 1}\ll\epsilon_{\ell 2}$, the electron does not play any important role in the $B$-anomalies, thus $\epsilon_{\ell 1}$ is not fixed by them if the latter limit is satisfied, as we will assume from now on.

Once the Left-handed mixings of $\mu$ and $\tau$ are fixed, the Right-handed ones can be adjusted to obtain the proper masses. Using Eq.~(\ref{eq-mixl}) one obtains: $\epsilon_{e3}\simeq 0.7\times10^{-2}/g_*$ and $\epsilon_{e2}\simeq 0.4\times10^{-3}/g_*$.

\subsection{Bounds}\label{sec-bounds}
One of the most stringent constraints on a $U_1$ leptoquark arises from LFU violation in $\tau$ decays. At one-loop $U_1$ modifies the $W$ coupling of the $\tau$, that is in agreement with the SM prediction at the per mil level. Following Ref.~\cite{Amhis:2016xyh} the violation of LFU can be parametrized in the ratio:
\begin{equation}
\left|\frac{g_\tau^W}{g_\mu^W}\right|=1.0000\pm 0.0014 \ .
\label{eq-tauLFUV1}
\end{equation}
One-loop radiative corrections can be estimated as:~\cite{Feruglio:2016gvd,Buttazzo:2017ixm}
\begin{equation}
\left|\frac{g_\tau^W}{g_\mu^W}\right|=1- 0.08 C^{3333}\ .
\label{eq-tauLFUV2}
\end{equation}
From Eqs.~(\ref{eq-tauLFUV1}) and~(\ref{eq-tauLFUV2}), we obtain $g_{L33}/m_{U_1}\lesssim 0.8/$TeV. This bound can be compared with the value required to fit $R_{D^{(*)}}$: $c_{23}g_{L33}/m_{U_1}\sim 1/$TeV. Although this puzzle seems to introduce some tension, a factor $c_{23}\sim 2-3$ is enough to satisfy both requirements. 

The large degree of compositeness of $\ell_{3}$ can induce large deviations in the couplings $Z\tau_L\bar \tau_L$ and $Z\nu\bar \nu$, that are in agreement with the SM at the per mil level also. In generic models with partial compositeness the correction to these couplings can be estimated as: $\delta g_\ell^Z/g_\ell^Z\sim\xi\epsilon_\ell^2$, with $\xi$ defined in Eq.~(24). However, in our model $\tau_L$ mixes with a resonance having $T^{3L}=T^{3R}$ and $(T^L)^2=(T^R)^2$, realizing a discrete $LR$ symmetry that protects $g_{\tau_L}^Z$~\cite{Agashe:2006at}. In this case the leading tree-level corrections are $\delta g_{\tau_L}^Z/g_{\tau_L}^Z\sim\xi\epsilon_{\ell 3}^2 (g/g_*)^2\sim {\rm few}\times 10^{-3}$ (see Ref.~\cite{Andres:2015oqa} for an explicit calculation in the case of the $b$-quark). Since it is not possible to protect $g_{\nu}^Z$ at the same time, this coupling gets a larger modification, requiring extra tuning of order $(g_*/g)^2$ to pass the constraints~\cite{ALEPH:2005ab}. We will show the numerical results performing a tree-level calculation in sec.~\ref{sub-boundsn}.

A similar situation holds for $Zb_L\bar b_L$. Given our choice for the embedding of the resonance mixing with the elementary $b_L$, the discrete $LR$ symmetry also protects $g_{b_L}^Z$, leading to corrections of order $10^{-3}$. In sec.~\ref{sub-boundsn} we will describe the numerical predictions in a three-site model.

As in any other model with anarchic partial compositeness from linear interactions, there are several quantities that push the compositeness scale to larger values: the neutron dipole moment that requires $f\gtrsim {\cal O}(5)$~TeV and $\epsilon_K$ in the Kaon-system that gives $m_*\gtrsim{\cal O}(10)$~TeV~\cite{Csaki:2008zd}. In the lepton sector the electron dipole moment and the flavor violating decay $\mu\to e\gamma$ give even stronger bounds: $f\gtrsim {\cal O}(20-40)$~TeV. There have been a few proposals for these problems, as the presence different scales for different flavors~\cite{Panico:2016ull}, the presence of naturally tiny bilinear interactions in anarchic scenarios~\cite{DaRold:2017xdm}, the existence of extended color symmetries in the SCFT~\cite{Bauer:2011ah,DaRold:2017dbr}, as well as the presence of flavor symmetries~\cite{Csaki:2008eh,Redi:2011zi}. The proposal of Ref.~\cite{DaRold:2017xdm}, where the lepton doublet and singlet of the first generation are elementary, as well as the first generation Right-handed up- and down-quarks, can be implemented straightforward in the present model. In this case, the most dangerous contributions to the aformentioned processes are suppressed, though transitions from operators like $(\bar s_Rd_L)^2$ and $(\bar d^i_L\gamma^\mu d^j_L)^2$ require $m_*\gtrsim 6-7$~TeV. An interesting alternative is proposed in Ref.~\cite{Frigerio:2018uwx}, where the authors consider a composite sector with $CP$ symmetry, as well as a flavor U(1)$^3$ symmetry in the composite ``leptonic'' sector. Either if the elementary-composite interactions respect U(1)$^3$, or if it is broken by the couplings $\lambda_\psi$, the constraints on $m_*$ are relaxed to $\lesssim 10$~TeV. Concerning the $B$-anomalies, Ref.~\cite{Frigerio:2018uwx} analyses two different cases: $\epsilon_{\ell i}\sim\epsilon_{ei}$, and $\epsilon_{\ell i}\sim\epsilon_{\ell j}$, showing that both scenarios can explain the anomalies. This proposal can be implemented straightforwardly in our model, by extending G to G$\times$U(1)$^3\times CP$. Although the composite fermion multiplets contain states with lepton and baryon number, the elementary fermions are not unified, in the sense that $q^i$ and $\ell^i$ interact with SCFT operators that have different charges under U(1)$^3$, thus in this scenario $\epsilon_{q i}$ and $\epsilon_{\ell i}$ are independent as required in our set-up (and similar for the Right-handed fermions).

As discussed in Ref.~\cite{Azatov:2018kzb}, bounds from $\bar B-B$ mixing combined with a solution to $R_{D^{(*)}}$ lead to $f\lesssim 0.7$~TeV. This condition introduces a tension with EW precision tests, that usually require, at least: $f\gtrsim 0.75$~TeV~\cite{Grojean:2013qca}, slightly increasing the amount of tuning of the model.

\subsection{Effective theory}\label{sec-effth}
An effective low energy theory, obtained after integration of the resonances of the SCFT, and containing the elementary fermions and gauge fields as well as the NGBs, can be built based only on symmetry principles. One of the main objects for this construction is the NGB unitary matrix:
\begin{equation}
U=e^{i\Pi/f}\ ,\qquad \Pi=\Pi^{\hat a} T^{\hat a} \ ,
\end{equation}
with $\Pi^{\hat a}$ the real NGB d.o.f., and $T^{\hat a}$ the broken generators of the coset SO(12)/SO(11). Under SO(12), $U$ transforms as: ${\cal G}U{\cal H}^\dagger$, with ${\cal G}\in$SO(12) and ${\cal H}\in$SO(11) being a function of ${\cal G}$ and $\Pi$: ${\cal H}({\cal G},\Pi)$.  In the fundamental and adjoint representations of SO(12), {\bf12} and {\bf66}, respectively, $U$ can be written as:
\begin{equation}
U= I+\frac{\sin(\rho/f)}{\rho}\Pi+\frac{\cos(\rho/f)-1}{\rho^2}\Pi^2  \ , \qquad \rho^2=\sum_{\hat a}(\Pi^{\hat a})^2
\label{eq-U}
\end{equation}

The kinetic term of the NGBs can be written in terms of the Maurer-Cartan form: $iU^\dagger D_\mu U=d_\mu^{\hat a}T^{\hat a}+e_\mu^aT^a$, where $D_\mu$ is the usual covariant derivative containing the SM gauge fields:
\begin{equation}\label{eq-lkinNGBs}
{\cal L}\supset \frac{f^2}{4}d_\mu^{\hat a}d^{\mu\hat a} \ ,
\end{equation}  
Assuming that only $H$ has a vev: $v$, Eq.~(\ref{eq-lkinNGBs}) generates a mass term for the electroweak (EW) gauge bosons, that is identical to the case of the MCHM based on the coset SO(5)/SO(4), namely:
\begin{equation}\label{eq-vSM}
v_{\rm SM}^2=(246{\rm GeV})^2=f^2\sin^2\left(\frac{v}{f}\right) \ .
\end{equation}
As usual in CHM we define:~\cite{Agashe:2004rs}
\begin{equation}\label{eq-xi}
\xi\equiv \frac{v_{\rm SM}^2}{f^2} \ .
\end{equation}

%\subsubsection{Fermions}
Let us focus now on the fermions. Although the elementary fields do not fill complete representations of SO(12), we find it useful to embed them in full SO(12) representations by adding non-dynamical fields, that must be put to zero in the end of the calculations. Following the discussions of sec.~\ref{sec-ferm}, we embed the elementary fermions in the representations of SO(12) shown in table~\ref{t-irreps}.

As usual in the CCWZ formalism~\cite{Coleman:1969sm,Callan:1969sn}, one can build SO(12)-invariants by dressing the fields with $U^\dagger$, and then forming with them SO(11)-invariants. Thanks to the transformation properties of $U$, these invariants are actually invariants of SO(12).~\footnote{As is well known, in the case of quadratic invariants depending on two fields that are in the same representation of G, there is a linear combination of invariants that is independent of the NGBs.}

Given a field $\psi$, that transforms with a representation of SO(12), we define $\psi_{\bf r}$ as the projection of $\psi$ to ${\bf r}$, with ${\bf r}$ a representation of SO(11). To quadratic order in the elementary fermions, the effective Lagrangian can be written as:
\begin{align}
&{\cal L}_{\rm eff}\supset 
\sum_fZ_f\bar\psi_f\pslash\psi_f
+\sum_{f,f'}\sum_{\bf r}(\bar\psi_fU)_{\bf r}\Pi_{ff'}^{\bf r}(U^\dagger\psi_f')_{\bf r} \nonumber \\
&f,f'=q,u,d,\ell,e
\label{eq-lf1}
\end{align}
The coefficient $Z_f$ stands for the elementary kinetic term, it will be taken to unity in numerical calculations. $\Pi_{ff'}$ are the form factors that codify the information arising from the integration of the resonances of the SCFT, they depend on momentum and on the microscopic parameters of the SCFT, but they not depend on the NGBs. Their precise form requires a model for the SCFT, as for example an extra-dimensional theory, or a discrete version of an extra dimension with a finite number of sites. In sec.~\ref{sec-3sites} we will show an explicit realization in terms of a three-site model.

If $f$ and $f'$ have the same chirality, $\Pi_{ff'}^{\bf r}$ is proportional to $\pslash$, in the following we will trade $\Pi_{ff'}^{\bf r}\to\pslash\Pi_{ff'}^{\bf r}$, factorizing that power of momentum from the form factors.

%\subsubsection{Gauge sector}
We also embed the elementary gauge fields in the adjoint representation of SO(12), by adding non-dynamical degrees of freedom. The effective Lagrangian at quadratic order in the elementary gauge fields $a_\mu$ is:
\begin{align}
&{\cal L}_{\rm eff}\supset 
\frac{1}{2} P_{\mu\nu} [-Z_g a^\mu p^2a^\nu
+\sum_{\bf r}(a^\mu U)_{\bf r}\Pi_{g}^{\bf r}(U^\dagger a^\nu)_{\bf r}] \ , 
\label{eq-lg1}
\end{align}
with $P_{\mu\nu}=\eta_{\mu\nu}-p_\mu p_\nu/p^2$, $Z_g=1/g_0^2$ and $\Pi_{g}^{\bf r}$ being the form factors that codify the SCFT dynamics after the integration of the spin one resonances, that are independent of the NGBs.

%\subsubsection{Effective theory in the vacuum of the Higgs vev}
It is useful to consider ${\cal L}_{\rm eff}$ with the NGBs evaluated in their vevs. Assuming that only the Higgs has a non-trivial vev $v$, and keeping just the dynamical fields corresponding to the SM, we obtain:
\begin{align}
{\cal L}_{\rm eff}\supset 
&\sum_{f=u,d,e,\nu}[\bar f_LM_ff_R+ {\rm h.c.}
+
\sum_{f=u,d,e,\nu}\sum_{X=L,R}\bar f_X\pslash(Z_{f_X}+\Pi_{f_X})f_X ] 
\nonumber \\
&+ \frac{1}{2}\sum_{a=g,w,b}\sum_j a^j_\mu (-Z_{a}p^2+\Pi_{a})a^j_\mu 
\ ,
\label{eq-lf2}
\end{align}
where the sum of the second line is over the dynamical gauge fields of the SM. We will call $g_{0s}$, $g_{0w}$ and $g_{0y}$ the elementary gauge couplings of SU(3)$_c$, SU(2)$_L$ and U(1)$_Y$, respectively, and: $Z_{a}=g_{0a}^{-2}$.
The field $\nu_R$ has not been written, it could be present or not, depending on the realization of neutrino masses. 

The functions $M_f$ and $\Pi_{f_X}$ can be computed by matching Eqs.~(\ref{eq-lf1}) and (\ref{eq-lf2}) in the background of the Higgs vev, they are given by:
\begin{align}
&M_f=\sum_{\bf r} j_{f}^{\bf r}\Pi_{qf}^{\bf r} \ , \qquad
\Pi_{f_R}=\sum_{\bf r} i_{f_R}^{\bf r}\Pi_{f}^{\bf r} \ , \qquad 
\Pi_{f_L}=\sum_{\bf r} i_{f_L}^{\bf r}\Pi_{q}^{\bf r} \ , \qquad
f=u,d \ ,
\nonumber\\
&M_e=\sum_{\bf r} j_{f}^{\bf r}\Pi_{\ell e}^{\bf r} \ , \qquad 
\Pi_{e_R}=\sum_{\bf r} i_{e_R}^{\bf r}\Pi_{e}^{\bf r} \ , \qquad 
\Pi_{f_L}=\sum_{\bf r} i_{f_L}^{\bf r}\Pi_{\ell}^{\bf r} \ , \qquad
f=e,\nu \ .
\label{eq-corf}
\end{align}

For the bosonic sector:
\begin{equation}
\Pi_{a}=i_a^{\bf r}\Pi_{g}^{\bf r} \ .
\label{eq-corg}
\end{equation}

The functions $i_{f}^{\bf r}, i_a^{\bf r}$ and $j_{f}^{\bf r}$ can be expressed in terms of trigonometric functions of $v/f$. Defining $s_v\equiv\sin(v/f)$ and $c_v\equiv \cos(v/f)$, we show them in table~\ref{t-inv1}. 
We only present the invariants involving fields with large degree of compositeness, that play an important role in the potential that determines the vev, the other invariants are straightforward to compute once the corresponding representations are built.
\begin{table}[ht]
\centering
\begin{tabular}{|c|c|c|c|c|c|c|c|c|c|}
\hline\rule{0mm}{5mm}
SO(12) & SO(11) & $i_{u_L}$ & $i_{d_L}$ & $i_{u_R}$ & $i_{\nu_L}$ & $i_{e_L}$ & $j_u$ & $i_{g}$ & $i_{w}$ \\[5pt]
\hline \rule{0mm}{5mm}
\multirow{2}{*}{\bf66}  & 
{\bf55}  & $1-s_v^2/2$ & $1$ & $s_v^2$ & $1-s_v^2/2$ & $1$ & $ic_v s_v/\sqrt{2}$ & 1 & $1-s_v^2/2$ \\[5pt]
\cline{2-10} \rule{0mm}{5mm}
 & {\bf11}  & $s_v^2/2$ & $0$ & $c_v^2$ & $s_v^2/2$ & $0$ & $-ic_v s_v/\sqrt{2}$ & 0 & $s_v^2/2$ \\[5pt]
\hline 
\end{tabular}
\caption{Invariants $i_{f}^{\bf r}$ and $j_{f}^{\bf r}$ of the kinetic and mass terms, in the background of the Higgs vev, with no vev for $\bar{S}_1$. We have used $s_v=\sin v/f$ and $c_v=\cos v/f$. We only show the invariants of the fields that have a non-negligible degree of compositeness.}
\label{t-inv1}
\end{table}

The spectrum of fermions of a given species, including the resonances that can be excited by the corresponding elementary fermion, can be obtained by computing the equations of motion of the dynamical elementary fermions present in the Lagrangian of Eq.~(\ref{eq-lf2}):
\begin{equation}
%{\rm zeroes}[
p^2(Z_{f_L}+\Pi_{f_L})(Z_{f_R}+\Pi_{f_R})-|M_f^2|=0 \ .
\label{eq-spec}
\end{equation}

The spectrum of the resonances that are not excited by the SM fermions, as the exotic ones (that have $Z_f=0$), can be obtained by computing the poles of Eq.~(\ref{eq-spec}) for each species of fermions.~\footnote{In some case there are also non-exotic resonances that are decoupled from the elementary SM fermions, thus their spectrum is given by the poles of the correlators.}

\subsection{Potential}
If the SCFT is considered in isolation, the Higgs, the singlet $\varphi$ and the scalar leptoquark are true NGBs. Since the interactions with the elementary sector explicitly break the global symmetry of the SCFT down to the SM group, at loop level a potential is generated, thus these scalars become pNGBs. The potential is dominated by the contributions of the elementary fields that have the largest couplings with the SCFT. In the present model, that role is played by $q_L$, $u_R$ and $\ell_L$ of the third generation, we will also consider the effect of the gluons $g^k$ and the weak fields $w^i$ on $V$, the effect of the other fields is sub-leading and it will not be taken into account.

At one-loop level, the Coleman-Weinberg potential can be written as:~\cite{DaRold:2018moy}
\begin{equation}\label{eq-VCW}
V= \frac{1}{2}\int \frac{d^4p}{(2\pi)^4}\left(-\log{\rm det}\frac{{\cal K}_f}{{\cal K}_f^0}+\log{\rm det}\frac{{\cal K}_a}{{\cal K}_a^0}\right)\ ,
\end{equation}
where ${\cal K}_f$ and ${\cal K}_g$ are the fermionic and bosonic matrices in the quadratic effective Lagrangian of elementary fields, Eq.~(\ref{eq-lf1}):
\begin{equation}\label{eq-LK}
{\cal L}_{\rm eff}\supset \bar f {\cal K}_ff + \frac{1}{2}a{\cal K}_aa\ , \qquad f^t=(u_L, d_L, \nu_L, e_L, u_R) \ ,\  a^t=(g^k,w^i) \ .
\end{equation}
For $f$ we have included only the fermions of the third generation giving the largest contribution to $V$, for $a$ we have included the elementary gauge fields of SU(3)$_c$: $g^k$, with $k=1,\dots 8$, and the ones of SU(2)$_L$: $w^i$, with $i=1,2,3$. We have neither shown spinor, nor vector, indices. The superindex 0 means that the NGBs are evaluated to zero, thus the denominators in the argument of the determinants just subtract a divergent term independent of the NGBs. Notice that, since the quarks have color indices, for our approximation ${\cal K}_f$ is a matrix of dimension eleven. The matrices ${\cal K}$ can be calculated by making use of Eqs.~(\ref{eq-U}) and (\ref{eq-lf1}). The specific form of the fermionic contribution depends on the embedding of the elementary fermions into SO(12), leaving freedom for model building. Since $\bar{S}_1$ is a singlet of SU(2)$_L$ and $H$ is a color singlet, at one-loop level the gluons contribute only to the potential of $\bar{S}_1$ and the $w$s only to the potential of the Higgs.

Since the dependence on the pNGBs is contained in the matrix $U$, $V$ is a complicated function of $\varphi$, $H$ and $\bar{S}_1$, with an infinite series of terms. In order to analyse the stability of the potential, we find it useful to perform an expansion of $V$ in powers of $\varphi$, $H$ and $\bar{S}_1$ to fourth order, obtaining:
\begin{equation}\label{eq-V4}
V \simeq \sum_{\Phi=H,\bar S_1,\varphi} [m_\Phi^2 |\Phi|^2 + \lambda_\Phi (|\Phi|^2)^2] + \lambda_{H\bar{S}_1} |H|^2|\bar{S}_1|^2 + \lambda_{H\varphi} |H|^2\varphi^2 + \lambda_{\varphi\bar{S}_1} \varphi^2|\bar{S}_1|^2 + \dots \ ,
\end{equation}
where the dots stand for higher order terms. The quadratic and quartic coefficients of Eq.~(\ref{eq-V4}) can be expressed as momentum integrals of combinations of the fermionic and bosonic form factors $\Pi_{ff'}^{\bf r}$ and $\Pi_g^{\bf r}$. Generically, the quadratic and quartic coefficients can be estimated to be of order: $m_\Phi^2\sim\epsilon_f^2m_*^4/(16\pi^2f^2)$ and $\lambda_\Phi\sim\epsilon_f^2m_*^4/(16\pi^2f^4)$. The absence of terms with an odd number of fields is guaranteed by the $P$-symmetry. Notice that in the absence of this symmetry, a term linear in $\varphi$ can be present, triggering a vev for $\varphi$.

For $m_H^2<0$, $m_{\bar{S}_1}^2,m_\varphi^2>0$ and suitable quartic couplings, $V$ is minimized by a non-trivial Higgs vev: $v^2=-m_H^2/\lambda_H$ and zero leptoquark and singlet vev. Using the estimates of the previous paragraph in the solution for the Higgs vev, for generic regions of the parameter space one obtains: $v\sim f$. As is well known, EWPT demands a separation between $v$ and $f$, leading to a tuning of order $\xi$, as usual in composite Higgs models.

For the embedding of table~\ref{t-irreps}, the quadratic coefficients are:
\begin{align}
m_H^2 = -\int \frac{d^4p}{(2\pi)^4} &\left(
2\frac{\Pi_\ell^{\bf55}-\Pi_\ell^{\bf11}}{Z_\ell+\Pi_\ell^{\bf55}}
+2N_c\frac{\Pi_q^{\bf55}-\Pi_q^{\bf11}}{Z_q+\Pi_q^{\bf55}}
+4N_c\frac{\Pi_u^{\bf11}-\Pi_u^{\bf55}}{Z_u+\Pi_u^{\bf11}}\right. \nonumber
\\
&\left.+2N_c\frac{|M_u^{\bf55}-M_u^{\bf11}|^2}{(Z_q+\Pi_q^{\bf55})(Z_u+\Pi_u^{\bf11})}
-\frac{9}{4}\ \frac{\Pi_g^{\bf11}-\Pi_g^{\bf55}}{-Z_w p^2+\Pi_g^{\bf55}}
\right) \ , 
\nonumber\\ 
m_{\bar{S}_1}^2 = -\int \frac{d^4p}{(2\pi)^4} &\left(4
\frac{\Pi_q^{\bf55}-\Pi_q^{\bf11}}{Z_q+\Pi_q^{\bf55}}
+10\frac{\Pi_u^{\bf11}-\Pi_u^{\bf55}}{Z_u+\Pi_u^{\bf11}}
-\frac{16}{3}\ \frac{\Pi_g^{\bf10}-\Pi_g^{\bf45}}{-Z_gp^2+\Pi_g^{\bf45}}
\right) \ ,
\nonumber\\ 
m_{\varphi}^2 = -\int \frac{d^4p}{(2\pi)^4} &4N_c
\frac{\Pi_u^{\bf11}-\Pi_u^{\bf55}}{Z_u+\Pi_u^{\bf11}}
\ ,
\label{eq-m2pi}
\end{align}
the last terms show the gauge contributions, that are independent of the fermion embedding. Having explicit expressions of the form factors it is possible to compute these coefficients.

We have also computed the quartic couplings, but we do not show them because they lead to too long expressions.

%\subsection{EWSB}
We consider now the potential that determines the Higgs vev, in the case of neither vev of $\bar{S}_1$, nor of $\varphi$. The dominant contributions in our model are given by:
\begin{align}
V\simeq \int \frac{d^4p}{(2\pi)^4}&\left\{\frac{9}{2}\log(-Z_wp^2+\Pi_w)-2\log[p^2(Z_\ell+\Pi_{e_L})]-2\log[p^2(Z_\ell+\Pi_{\nu_L})]\right.\ ,\nonumber
\\
&\left.-2N_c\log[p^2(Z_q+\Pi_{d_L})]-2N_c\log[p^2(Z_{u_L}+\Pi_{u_L})(Z_{u_R}+\Pi_{u_R})-|M_u|^2]\right\} \ ,
\label{eq-VCWv}
\end{align}
%\left(1+\frac{s_v^2}{2}\frac{\Pi_g^{\bf10}-\Pi_g^{\bf45}}{-Z_w+\Pi_g^{\bf45}}\right)
where the form factors can be obtained by making use of the Eqs.~(\ref{eq-corf}) and (\ref{eq-corg}), as well as table~\ref{t-inv1}.

In the next section we will show an effective description of the resonances of the SCFT that allows to model the form factors. In sec.~\ref{sec-ewsb} we will show numerical results for regions of the parameter space where the one-loop potential breaks the EW symmetry, preserving SU(3)$_c\times$U(1)$_{\rm em}$. We will also show the predictions for the masses of the pNGBs.

\section{A three-site model}\label{sec-3sites}
\begin{figure}[t]
\centering
\includegraphics[width=0.7\textwidth]{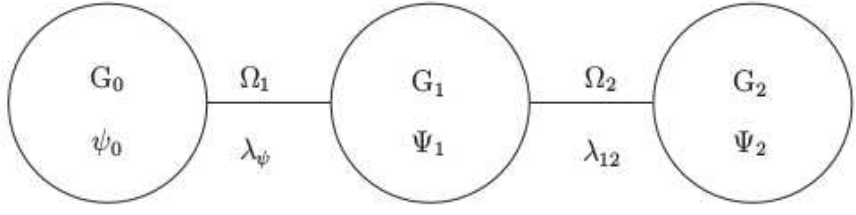}
\caption{Moose diagram of the three-site theory.}
\label{fig-moose}
\end{figure}

We consider an effective description of the SCFT dynamics and its interactions with the elementary SM fermions and gauge bosons in terms of a discrete composite model. The minimal set-up that allows to obtain the proper leptoquark interactions, as well as a finite one-loop potential is a three-site theory~\cite{Panico:2011pw}. We show a moose diagram of such a theory in Fig.~\ref{fig-moose} (see Ap.~\ref{ap-extradim} for a model in 5-dimensions). Site-0 contains the elementary fermions and gauge fields of the SM, that will be denoted with small letters, whereas the other two sites contain the lightest set of resonances of the SCFT, whose fields will be denoted with capital letters and subindices indicating the site. Site-1 has a local symmetry G$_1$=SO(12), as well as massive Dirac fermions transforming with irreducible representations of G$_1$. Site-2 has massive Dirac fermions transforming with irreducible representations of a global symmetry G$_2$=SO(12), however only a subgroup H$_2$=SO(11) is gauged on this site. Two nearest sites $j$ and $j+1$ are connected by $\sigma$-models based on the coset G$_j\times$G$_{j+1}/$G$_{j+(j+1)}$, with a field $\Omega_j=e^{i\Pi_j/f_j}$ transforming linearly under G$_j\times$G$_{j+1}$ and parametrizing the coset. 

As described in sec.~\ref{sec-effth}, we find it useful to add non-dynamical elementary degrees of freedom that allow to embed the elementary fields in full multiplets of SO(12). Using these embeddings it is straightforward to write invariants, the spurion fields are set to zero in the end of the calculation. We will call $a_\mu$ to the embedding of the gauge fields into SO(12), and $\psi_f$ to the embeddings of the elementary fermions, with $f=q,u,d,\ell,e$, the later ones are defined in table~\ref{t-irreps}. 

The Lagrangian of the bosonic sector is:
\begin{equation}
{\cal L}_{\rm b}=-\frac{1}{4g_0^2}f^a_{\mu\nu}f^{a\mu\nu}+\sum_{j=1,2}\left[-\frac{1}{4g_j^2}F^a_{j\mu\nu}F_j^{a\mu\nu}+\frac{f_j^2}{4}{\rm tr}(|D_\mu \Omega_j|^2)\right] \ ,
\label{eq-l3b}
\end{equation}
with $f_{\mu\nu}$ and $F_{j\mu\nu}$ the field strength on site-0 and $j$, and $D_\mu \Omega_j=\partial_\mu \Omega_j+iA_{j-1\mu}\Omega_j-i\Omega_jA_{j\mu}$ ($a_\mu\equiv A_{0\mu}$). $g_j$ are the gauge couplings on site-$j$, with $g_{\rm SM}\ll g_1,g_2\ll 4\pi$. Matching the coupling of the unbroken diagonal group leads to: $g^{-2}_{\rm SM}=g_{0}^{-2}+g_{1}^{-2}+g_{2}^{-2}$, fixing $g_0$, whose size can be estimated as $g_0\sim g_{\rm SM}$ for $g_1,g_2\gg g_{\rm SM}$. In fact a subindex in the coupling at site-0, as well as in $g_{SM}$, must be understood, distinguishing the different factors of the SM gauge symmetry.

The $\sigma$-model fields provide 132 NGBs, in the unitary gauge 121 NGBs become the longitudinal degrees of freedom of the 66 spin one resonances at site-1 and the 55 resonances at site-2, that become massive. 11 NGBs remain in the spectrum, they correspond to $H$, $\bar{S}_1$ and $\varphi$. By going to the unitary gauge, one can obtain the decay constant of the physical NGBs: $f$, as:
\begin{equation}
\frac{1}{f^2}=\frac{1}{f_1^2}+\frac{1}{f_2^2} \ .
\label{eq-matchf}
\end{equation}

The Lagrangian of the fermionic sector is:
\begin{align}
{\cal L}_{\rm f}=&\sum_fi\bar \psi_f\Dslash \psi_f +\bar\Psi_1^{\bf R}(i\Dslash-m_1^{\bf R})\Psi_1^{\bf R}+i\bar\Psi_2^{\bf R}\Dslash\Psi_2^{\bf R}
+\sum_{\bf r}m_{2,\bf r}^{\bf R}\bar\Psi_{2,\bf r}^{\bf R}\Psi_{2,\bf r}^{\bf R}
\nonumber\\
&+f_1\sum_f \lambda_f^{\bf R} \ \bar \psi_f\Omega_1\Psi_1^{\bf R}
+\lambda_{1,2}^{\bf R}f_2\ \bar\Psi_1^{\bf R}\Omega_2\Psi_2^{\bf R} + {\rm h.c.} \ , \nonumber\\
f=&q,u,d,\ell,e \ ,
\label{eq-l3f}
\end{align}
where a generation index is understood. The superindex ${\bf R}$ labels the SO(12) representation of the corresponding fermion. The first line of Eq.~(\ref{eq-l3f}) contains the kinetic terms of the elementary fermions $f$, as well as the kinetic and mass terms of the fermions on sites 1 and 2. The last term of the first line contains masses for the fermions on site-2, with $\bf r$ being the representations obtained after the decomposition of $\Psi_2$ under SO(11), see Eq.~(\ref{eq-embed2}) for $\Psi_2\sim${\bf66} or {\bf220}. These terms, that give different masses to each SO(11) multiplet of fermions, are allowed because on site-2 only SO(11) is gauged. The second line contains the mixing between fermions located on different sites, these terms are gauge invariant thanks to the transformation properties of the matrices of NGBs. The masses are taken as $m_1\sim g_1 f_1$ and $m_{2,\bf r}\sim g_2 f_2$, such that fermionic and bosonic resonances have masses of the same size. The mixing parameters $\lambda_f$ are dimensionless numbers that, as discussed below Eq.~(\ref{eq-linearint}), can span a hierarchy of values, being very small for the fermions of first and second generations, as well as $b_R$ and $\tau_R$, and ${\cal O}(g_1)$ for the other fermions of the third generation.

Following the discussion of sec.~\ref{sec-ferm}, one would have to include two massive Dirac fermions per generation in each composite site, one in the representation ${\bf66}$ and one in ${\bf220}$. The elementary fermions $q$, $\ell$ and $u$ mix with $\Psi^{\bf66}_1$, whereas $d$ and $e$ mix with $\Psi^{\bf 220}_1$.
From Eqs.~(\ref{eq-mixq}) and (\ref{eq-mixl}) and the discussions of sec.~\ref{sec-Ba}, we obtained that $q$, $u$ and $\ell$ of the third generation have a large degree of compositeness, whereas the mixing $\lambda_d$ and $\lambda_e$ must be suppressed to reproduce the bottom and tau masses. Given that the mixings with $\Psi^{\bf 220}_1$ are very suppressed, to simplify our analysis of the potential, as well as the analysis of $q$ and $\ell$ couplings with leptoquarks, from now on we will consider that there is just one resonance in each site: $\Psi_j^{\bf66}$, and we will neglect the effects of $\Psi_j^{\bf220}$.~\footnote{Even if $\lambda_d=\lambda_e=0$, $\Psi_2^{\bf220}$ can mix with $\Psi_2^{\bf66}$ through a mass term that is SO(11)-invariant: $\hat m_2\bar\Psi_{2L,\bf55}^{\bf66}\Psi_{2R,\bf55}^{\bf220}+\hat m'_2\bar\Psi_{2R,\bf55}^{\bf66}\Psi_{2L,\bf55}^{\bf220}$, distorting the spectrum of fermions. We are neglecting these effects also.} Thus Eq.~(\ref{eq-l3f}) simplifies to: 
\begin{align}
{\cal L}_{\rm f}= &\bar\Psi_1^{\bf66}(i\Dslash-m_1)\Psi_1^{\bf66}+i\bar\Psi_2^{\bf66}\Dslash\Psi_2^{\bf66}
+\sum_{\bf r=55,11}m_{2,\bf r}\bar\Psi_{2,\bf r}^{\bf66}\Psi_{2,\bf r}^{\bf66}+\lambda_{1,2}f_2\ \bar\Psi_1^{\bf66}\Omega_2\Psi_2^{\bf66}
\nonumber\\
&+f_1\sum_{f=q,\ell,u}\lambda_f(\bar\psi_f\Omega_1\Psi_1^{\bf66}+i\bar\psi_f\Dslash\psi_f) + {\rm h.c.} \ .
\label{eq-Lf66}
\end{align}

The mass matrices obtained from Eqs.~(\ref{eq-l3b}) and (\ref{eq-Lf66}) are shown in Ap.~\ref{ap-mm}.

The states at sites 2 and 3 give an effective description of the lightest level of resonances of the SCFT. To gain some insight into the dynamics of the SCFT, it is useful to study first the spectrum neglecting the mixing with the elementary fields as well as the contributions from the Higgs vev. Let us start with the spin one states, there is one multiplet in the representation {\bf11} of SO(11), located on site-1, with mass $g_1\sqrt{(f_1^2+f_2^2)/2}$, that does not mix with any other state in this limit. There are two multiplets in the representation {\bf55} of SO(11), that are mixed by $f_2$ as shown in Eq.~(\ref{eq-l3b}), after diagonalization of this mixing, for $f_1=f_2$ and $g_1=g_2$ the mass eigenstates have masses: $\simeq g_1 f_1/2$ and $\simeq g_1 f_1$. Turning-on $g_0$ the elementary and composite states are mixed. After this mixing, and before EWSB, one obtains a set of massless and partially composite fields that are in one to one correspondence with the gauge bosons of the SM. 
The degree of compositeness of the massless states: $\epsilon_g$, is defined as
\begin{equation}
\epsilon_g^2= 1-\frac{g_1^2g_2^2}{g_0^2g_1^2+g_0^2g_2^2+g_1^2g_2^2}\simeq g_0^2\left(\frac{1}{g_1^2}+\frac{1}{g_2^2}\right) + {\cal O}\left(\frac{g_0}{g_i}\right)^4 \ ,
\label{eq-alphag}
\end{equation}
it can be obtained by calculation of the massless eigenstate of the mass matrix before EWSB. $\epsilon_g$ can be defined as $\epsilon_g=\sqrt{1-x_{g,\rm el}^2}$, where $x_{\rm el}$ is the projection of the massless eigenstate onto the elementary one.

For the fermions one can proceed in an analogous way. Taking the limit of $\lambda_f=0$, the masses of the fermionic resonances depend on the masses at each site, as well as on the mixings between sites:  $\lambda_{1,2}$. There are two multiplets in the {\bf55} that split in two levels, as well as two multiplets in the {\bf11} that split in two levels, with masses:
\begin{equation}
m_{\bf r}^{(\pm)}=\frac{1}{2}\{m_1+m_{2,\bf r}\pm[(m_1-m_{2,\bf r})^2+4\lambda_{1,2}^2f_2^2]\} \ , \qquad {\bf r=55,11} \ .
\label{eq-specf}
\end{equation}
Since we will take $\lambda_{1,2}\sim g_1\sim g_2$, and all the fermionic mass parameters of order $g_j f_j$, the masses of the fermionic resonances, before mixing with the elementary states, are of the same size as the masses of the spin one resonances. 

One can also perform a biunitary diagonalization of the fermionic mass matrices before EWSB. Considering just one generation, we define the degree of compositeness as:
\begin{align}
&\epsilon_q=\lambda_q^2\frac{\lambda_{1,2}^2f_2^2+m_{2,\bf55}^2}{(m_1m_{2,\bf55}-\lambda_{1,2}^2f_2^2)^2+\lambda_q^2(m_{2,\bf55}^2+\lambda_{1,2}^2f_2^2)^2}
\ , \nonumber\\
&\epsilon_u=\epsilon_q (\lambda_q\to\lambda_u,m_{2,\bf55}\to m_{2,\bf11} )
\ , \nonumber\\
&\epsilon_\ell=\epsilon_q (\lambda_q\to\lambda_\ell)\ .
\label{eq-alphaf}
\end{align}
The mixing with the elementary states also shifts the masses of the resonances, leading to what is usually known as light custodians when the mixing is large.

In the next sections we will show numerical predictions for the spectrum and couplings. We will take into account all the mixings for these predictions, including those induced by the Higgs vev.

Integrating the resonances at sites 2 and 3, it is possible to obtain explicit expressions for the form factors of the low energy effective theory of sec.~\ref{sec-effth}. For the fermions we obtain:
\begin{align}
&\Pi_{qq}^{\bf r}=S(\lambda_q,m_{2,\bf r}) \ ,
\qquad
\Pi_{uu}^{\bf r}=S(\lambda_u,m_{2,\bf r}) \ ,
\nonumber\\
&\Pi_{\ell\ell}^{\bf r}=S(\lambda_\ell,m_{2,\bf r}) \ ,
\qquad
\Pi_{qu}^{\bf r}=M(\lambda_q,\lambda_u,m_{2,\bf r}) \ ,
\end{align}
with $S$ and $M$ defined as:
\begin{align}
&S(\lambda,m_{2,\bf r})=\lambda^2f_1^2(p^2-\lambda_{1,2}^2f_2^2-m_{2,\bf r}^2)/d \ ,
\nonumber\\
&M(\lambda,\lambda',m_{2,\bf r})=\lambda\lambda'f_1^2[(-p^2+m_{2,\bf r}^2)m_1-\lambda_{1,2}^2f_2^2m_{2,\bf r}]/d \ ,
\nonumber\\
&d=[\lambda_{1,2}^2f_2^2+(m_1-p)(p-m_{2,\bf r})][-\lambda_{1,2}^2f_2^2+(m_1+p)(p+m_{2,\bf r})] \ .
\end{align}
For the gauge fields we get:
\begin{align}
&\Pi_g^{\bf55}=\frac{p^2f_1^2[2p^2-f_2^2(g_1^2+g_2^2)]}{4p^4-2p^2g_1^2(f_1^2+f_2^2)+g_2^2f_2^2(-2p^2+g_1^2f_1^2)} \ ,
\nonumber\\
&\Pi_g^{\bf11}=\frac{f_1^2(2p^2-f_2^2g_1^2)}{4p^2-2g_1^2(f_1^2+f_2^2)} \ .
\end{align}

%%%%%%%%%%%%%%%%%%%%%%%%%%%%%%%%%%%%%%%%%%%%%%%%%%%%%%%%%%%%%%%%
\section{Phenomenology}\label{sec-pheno}
In this section we study the phenomenology of the composite GUT, with the lowest level of resonances effectively described by the three-site model of the previous section. We compute first the spectrum of pNGBs, looking for regions of the parameter space where the SM states have the proper masses, and showing the predictions for the masses of $\varphi$, $\bar S_1$ and the spin one states $U_1$, as well as the lightest $Z'$. After that we study the corrections to the $Z$ couplings of $\tau_L$ and $b_L$ at tree level, showing that, thanks to the $P_{LR}$-symmetry, the corrections are of order $\sim {\rm few}\times 0.1\%$ for $\xi\sim 0.1$. After that, we study the corrections to $R_{D^{(*)}}$ and show that, for large degree of compositeness of $\tau_L$, it is possible to be within $1\sigma$ of the experimental average value. In this case the correction to $W\tau$ coupling saturates the bounds. Finally, we discuss the phenomenology of the pNGBs.

\subsection{EW symmetry breaking and spectrum of resonances}\label{sec-ewsb}
We  have performed a scan of the parameter space of the three-site model, calculating the pNGB potential and the Higgs vev. For this, we let most of the parameters to vary randomly, with a number of constrains. As we wish to maintain perturbativity, at each site dimensionful couplings should not exceed $4 \pi f_n$. At sites 1 and 2, gauge couplings were chosen in the interval [2.1,3.6], and the couplings at the elementary site where adjusted to match the value of the SM couplings, as discussed below Eq.~(\ref{eq-l3b}). The scan was also optimized to select the elementary-composite couplings such as to have the fermionic degrees of compositeness larger than 0.5, as the points consistent with phenomenology would be at larger mixings. For each point we calculated the one-loop potential at all orders in the Higgs vev, obtaining the value of $v$ in the cases with EWSB. We discarded all the points with maximal breaking, as well as those without breaking. For each point kept, we obtained the Higgs mass by calculating the curvature around the minimum. The top mass is calculated as well, by biunitary diagonalization of the up quark mass matrix in presence of vev. Last, by rescaling all the dimensionful parameters, we fixed: $\sqrt{\xi}f=246\ {\rm GeV}$, as indicated in the matching of Eq.~(\ref{eq-vSM}).
%\subsection*{Benchmark window}
\begin{figure}[t]
\centering
  \includegraphics[width =0.7\textwidth]{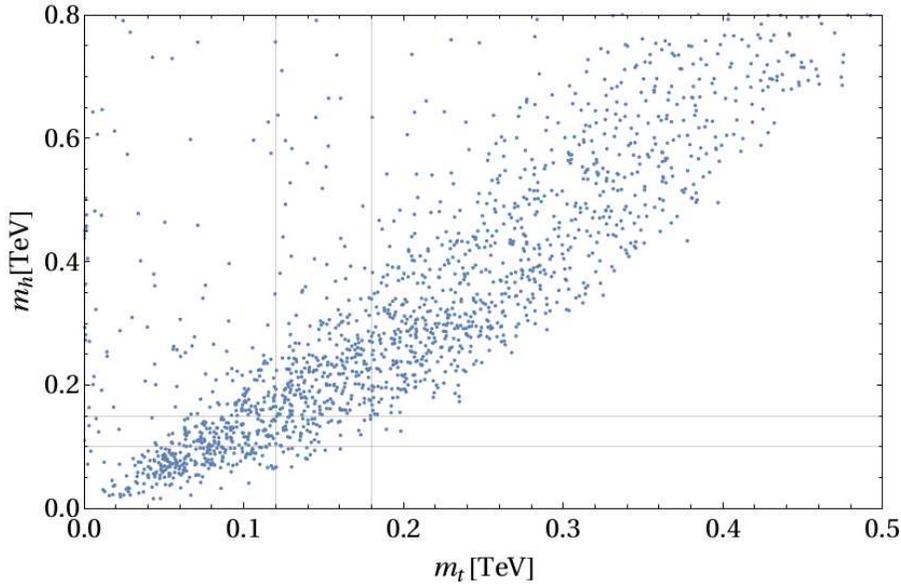}
  \caption{Scatter plot for top mass versus Higgs mass. Only points with $f \geq 0.7$ TeV were selected for this plot. In grey we highlight the limits for the benchmark window, defined by  $m_t \in \left[0.12,0.18\right]$TeV and $m_h \in \left[0.1,0.15\right]$ TeV. }
\label{fig-mtmh}
\end{figure}

We defined a benchmark window, selecting phenomenologically viable points, as $f \geq 0.7$~TeV, $m_t \in \left[120,180\right]$~GeV and $m_h \in \left[100,150\right]$~GeV, a smaller window requires more time of CPU running, with almost no impact in the phenomenology that we want to study. The $W$ mass was fixed to its experimental value by the rescaling described in the previous paragraph, whereas the $Z$ mass is related with the $W$ one by custodial symmetry. In Fig.~\ref{fig-mtmh} we show a scatter plot of $m_t$ versus $m_h$ obtained with the random scan, selecting points with $f>700\ {\rm GeV}$. The random points contain the phenomenologically interesting region of the SM, although the model prefers a ratio $m_h/m_t$ slightly larger than the experimental value.

\begin{figure}[h]
\centering
\includegraphics[width=0.8\textwidth ]{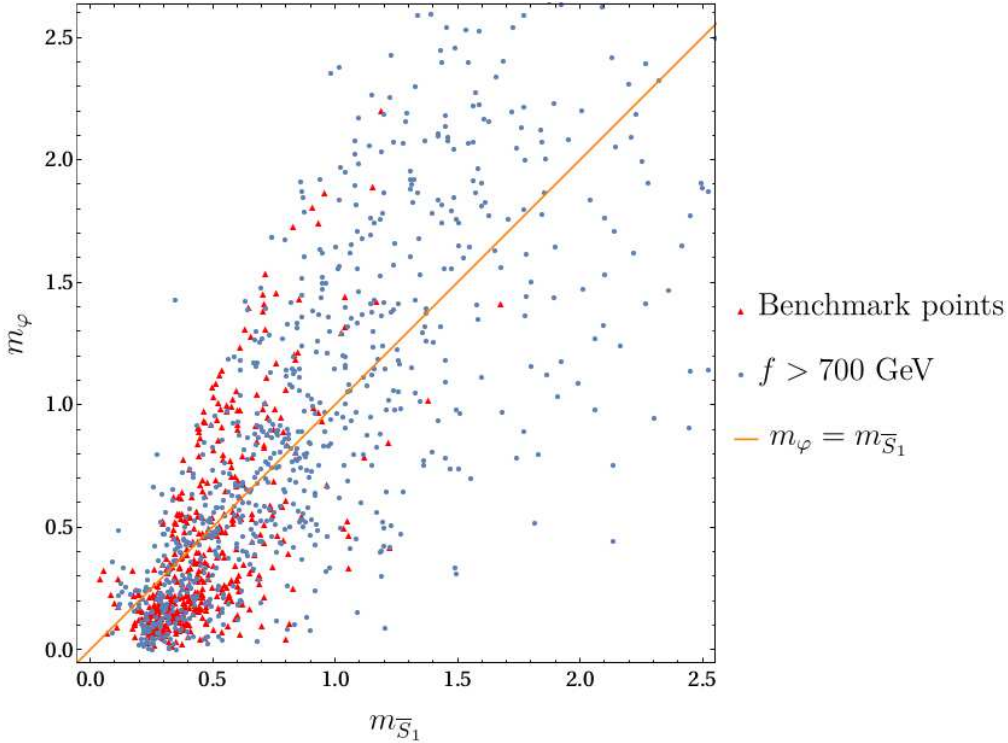} % =\textwidth works better
\caption{Scatter plot of the $\varphi$ particle mass as a function of the  mass of the scalar LQ. Plotted in blue are points with f $>$ 700 GeV and in red the ones inside the benchmark window. Also included is the line $m_\varphi = m_{\bar S_1}$.}
\label{scalmass}
\end{figure}

We have also calculated the masses of the other scalar states by taking into account both: the gauge and the fermion contribution to the potential, as shown in Eq.~(\ref{eq-m2pi}). The gauge contribution is always positive, whereas the fermionic one can be either positive or negative. In Fig. \ref{scalmass} we show both masses plotted one against the other, along with the line $m_\varphi = m_{\bar S_1}$, we only show the results with both masses positive. Red triangles correspond to points within the benchmark region, whereas blue circles are for points that lie outside that region. Just by counting red triangles we obtain that there are more points with $m_{\bar S_1}>m_\varphi$, than the other way around. For masses larger than 1~TeV it seems that there are more red triangles with $m_{\bar S_1}>m_\varphi$, however, since the blue circles do not show that pattern, we can not be sure if this a fluctuation due to the somewhat small number of points lying in the benchmark region, or if it is a reliable tendency. The masses of these states are in the range 0.1-2.5~TeV, with a larger density of points in the interval 100-700~GeV. 

\begin{figure}[h]
\centering
\includegraphics[width=\textwidth ]{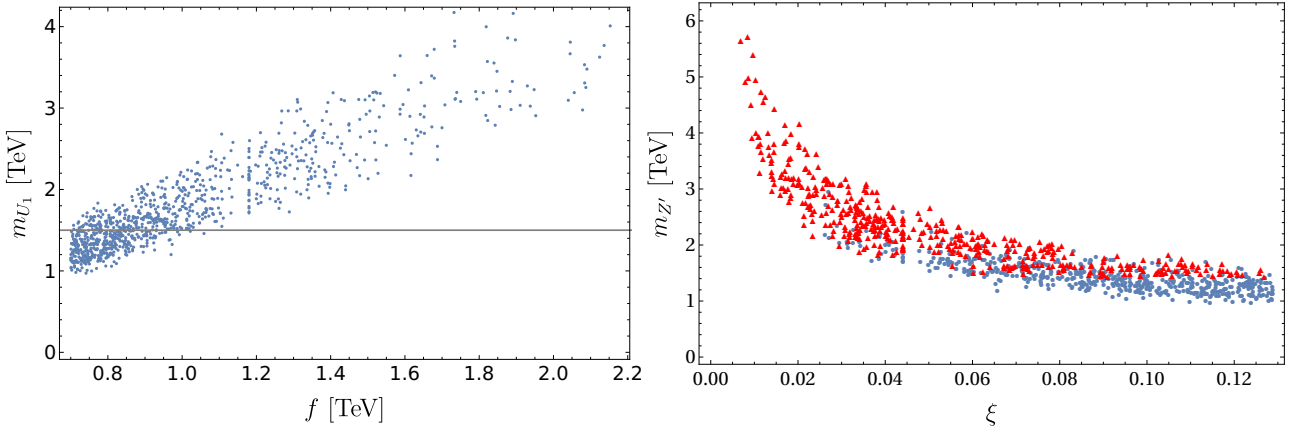} 
\caption{On the left panel we show a scatter plot of mass of the vector leptoquark $U_1$, as a function of the Higgs decay constant $f$, for the benchmark region. We observe the predicted linear dependence with $f$. Also pictured is the line at 1.5 TeV, often cited as a bound for its mass in direct searches~\cite{Biswas:2018snp}. On the right panel we show a scatter plot of mass for the first $Z'$ resonance, as a function of $\xi = v_{\rm SM}^2/f^2$, for points on the benchmark window. Red triangles pass the experimental constraints, whereas blue circles do not.}
\label{lqzprimemass}
\end{figure}

For the vector leptoquark $U_1$ we calculated its mass and its coupling to $b_L$ and $\tau_L$. This was also done by diagonalizing the fermion and boson mass matrices, and by writing the Lagrangian in terms of physical degrees of freedom. As there are three $U_1$ states, we obtained three masses and three different couplings. For the fermion embeddings that we have chosen, the parity $P$ implies that one of these states does not couple to the physical fermions, and as such it does not contribute to the couplings $R_{D^{(*)}}$. In Fig.~\ref{lqzprimemass} we present the mass of the lightest $U_1$ state, as a function of the decay constant of the pNGBs, for the benchmark region. The dependence with $f$ can be understood from the discussion of the spectrum above Eq.~(\ref{eq-specf}). We obtain that, for the benchmark region, $m_{U_1}\sim 1-4$~TeV, with a larger density of points near $1-2$~TeV. In the next section we will show the ratio $g_{U_1^{(n)}}/m_{U_1^{(n)}}$.

By calculating the eigenvalues of the mass matrix of the spin one neutral states, we find the spectrum of neutral vector resonances. Besides the SM $Z$ boson, the next 5 states are of two kinds. The lightest 4 are actually degenerate in mass, and do not mix with the elementary $Z$, while the 5th one is around 5\% heavier, and does mix with the elementary $Z$. As the first generation of quarks and leptons have a very low degree of compositeness, in the approximation in which they are fully elementary, their couplings to these fully composite $Z'$ is zero. As such, their generation in $p p$ collisions is highly suppressed, as the parton distribution functions of the 2nd and 3rd generations in the proton are suppressed. A process that can put bounds on these $Z'$ with couplings mostly to the third generation is four tops, which constrains the size of 4 quark operators that can be generated as a $Z'$ exchange~\cite{Sirunyan:2019nxl}. However, all points in the benchmark window pass these constraints, as they are not too restrictive. In that case, points were only accepted or rejected according to the lightest $Z'$ that does couple to first generation, following experimental bounds on these heavy vector bosons into either leptons, light jets, bottom and top quarks~\cite{Aaboud:2017sjh,Aaboud:2017fgj,Aaboud:2018tqo,Sirunyan:2018ryr,Sirunyan:2019vgt,CMS:2019tbu}. In Fig.~\ref{lqzprimemass} we present $m_Z'$ as a function of $\xi$, for points inside the phenomenological window. Using that $m_{Z'}$ scales with $f$, as well as Eq.~(\ref{eq-vSM}), one can understand the dependence of $m_{Z'}$ with $\xi$. We see that $m_{Z'} \gtrsim 2$ TeV for values of $\xi \lesssim 0.1$. The distinction between points that are ruled out by $Z'$ detection and those that pass the experimental constraints can be seen by the marker, where red triangles are points that pass, and blue circles are those that do not. 

\subsection{Bounds}\label{sub-boundsn}
We have computed the coupling of $Z$ to $\tau_L$, $\nu$ and $b_L$ in our three-site model. By rewriting the three-site interaction Lagrangian in terms of the physical states, one finds the value of the couplings $g_\tau^Z$, $g_\nu^Z$ and $g_b^Z$. 

\begin{figure}[h]
\centering
\includegraphics[width=0.47\textwidth]{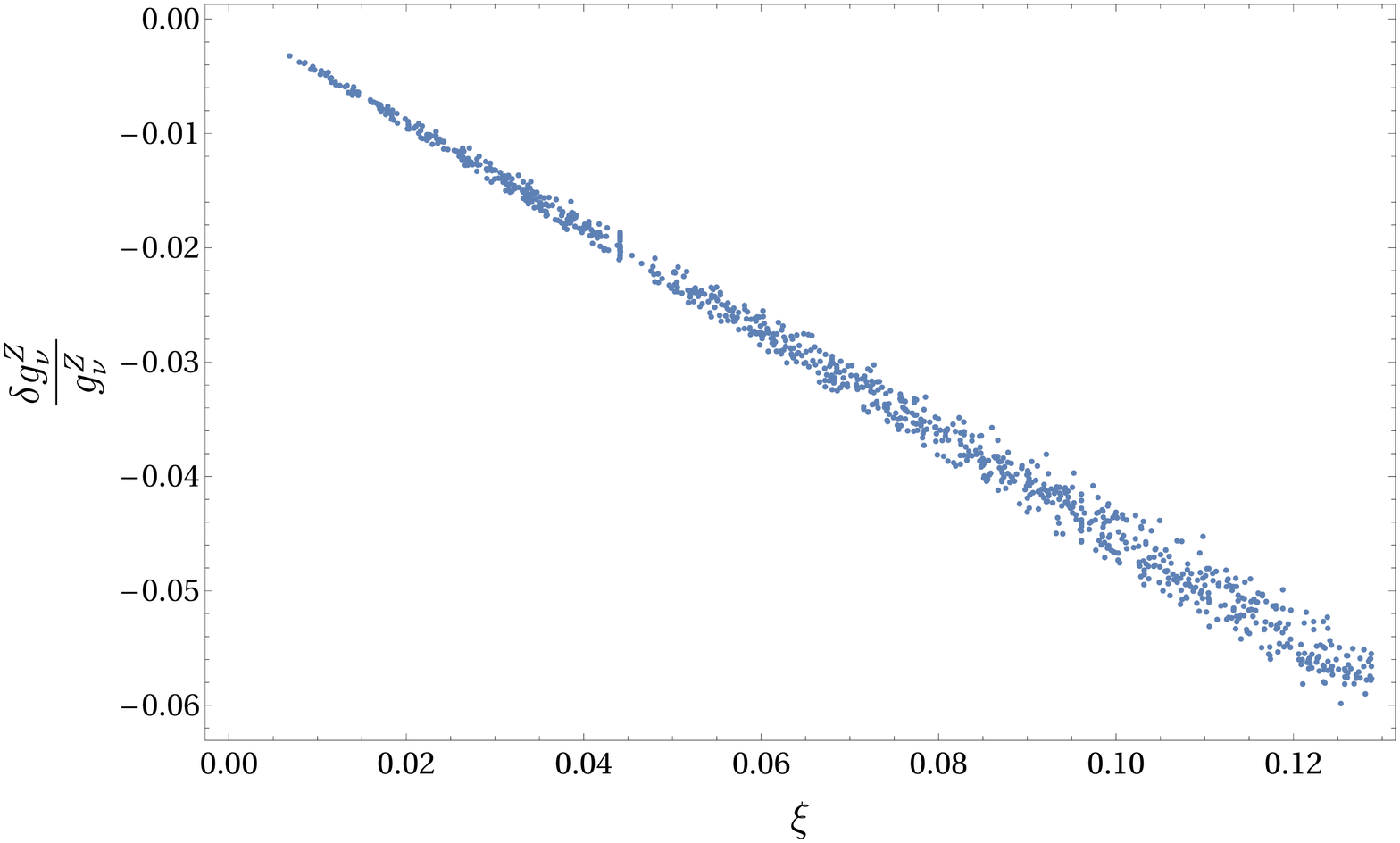}
\hskip0.4cm
\includegraphics[width=0.47\textwidth]{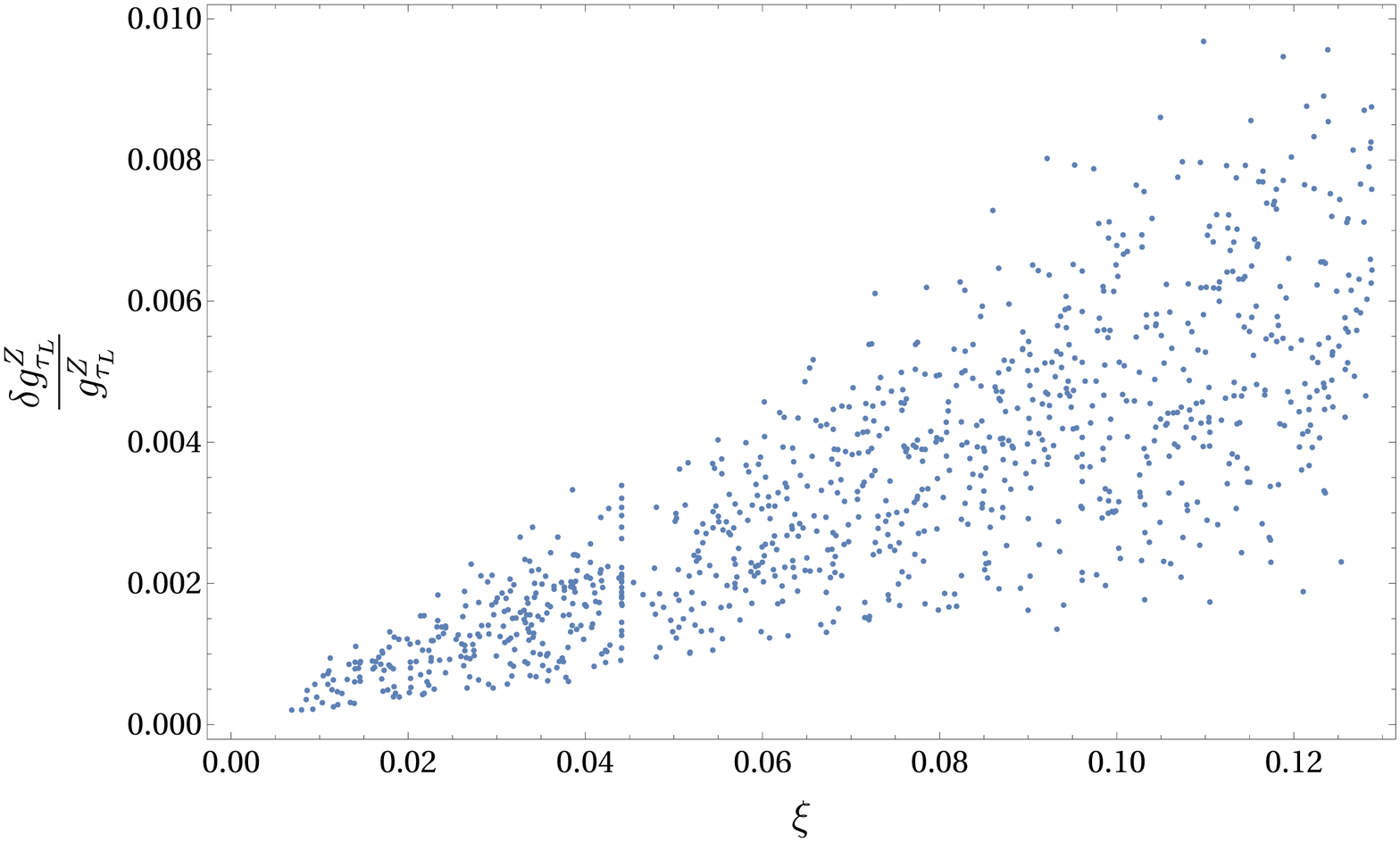}
\caption{Relative correction of the couplings $Z\nu\bar\nu$ on the left panel, and $Z\tau_L \bar\tau_L$ on the right panel, as a function of $\xi$, for points in the benchmark region.}
\label{ztautau}
\end{figure}

The coupling $g_\tau^Z$ has been measured with an accuracy of order few per mil~\cite{ALEPH:2005ab}. In the right panel of Fig.~\ref{ztautau} we present the relative difference of $g_{\tau_L}^Z$ coupling, as a function of $\xi$. As expected from the estimates of sec.~\ref{sec-bounds}, it scales linearly with $\xi$, the dispersion arising from the dependence of the coupling on $\epsilon_{\ell}$ and $\epsilon_g$. We find that the relative corrections to the coupling can reach values below 2-5 per mil from $\xi \lesssim 0.1$ onwards. The coupling $g_\nu^Z$ has also been measured at the per mil level~\cite{ALEPH:2005ab}, but in this case, since there is no symmetry protection, the corrections are a factor $1/\epsilon_g^2$ larger than for the charged lepton. As can be seen in the left panel of Fig.~\ref{ztautau}, the bounds require $\xi \lesssim 0.02$, increasing the amount of tuning.

We omit the same graph for the coupling of $Z$ to bottom quark, since the values and distribution of this coupling are similar. 

As described in sec.~\ref{sec-bounds}, LFU violation in $W$ couplings give constraints on $C^{3333}$. We have computed this Wilson coefficient for the benchmark points. In the next section we will distinguish the points that pass these constraints form those that do not, showing that a large portion of the former can also explain $R_{D^{(*)}}$.

\subsection{Predictions for $R_{D^{(*)}}$}\label{sec-RDn}
\begin{figure}[h]
\includegraphics[width=0.8\textwidth]{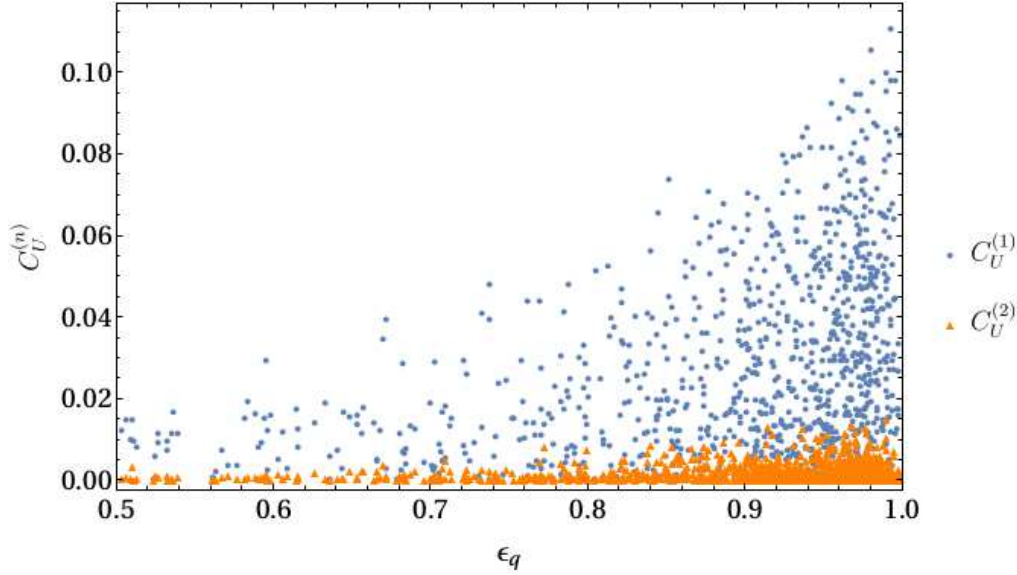}
\centering
\caption{Scatter plot of $C_U^{(n)} = 1/2(vg_{L33}^{(n)}/m_{U_1}^{(n)})^2$ for points inside the benchmark window, as a function of the degree of compositeness of the quark doublet of the third generation: $\epsilon_q$.}
\label{cus}
\end{figure}

The contribution of the $U_1$ leptoquarks to $R_{D^{(*)}}$ is given in Eq.~(\ref{eq-RD1}). Making use of the flavor structure arising from partial compositeness, $C^{3233}$ can be estimated as:
\begin{equation}\label{eq-c3233}
C^{3233}\left(1-\frac{V_{tb}^*}{V_{ts}^*}\frac{g_{L23}}{g_{L33}}\right)
\sim C^{3333}\lambda_C^{2}\frac{c_{23}}{c_{33}}\left(1+\frac{c_{23}}{c_{33}}\right) \ ,
%\sim\lambda_C^{2}\sum_nC_U^{(n)}
\end{equation}
with
\begin{equation}\label{eq-c3333}
C^{3333}=\sum_nC_U^{(n)} \ ,\qquad
C_U^{(n)} \equiv \frac{1}{2} \left(\frac{v g_{L33}^{(n)}}{m_{U_1}^{(n)}}\right)^2\ .
\end{equation}
Thus, up to a factor of ${\cal O}(1)$, arising from the last two factors of the r.h.s. of Eq.~(\ref{eq-c3233}), we can obtain $C^{3233}$ by knowing $g_{L33}^{(n)}/m_{U_1^{(n)}}$.
In Fig.~\ref{cus} we present the coefficients $C_U^{(n)}$ of both $U_1$ states having nonzero couplings to $b$ and $\tau$, as a function of the bottom quark mixing $\epsilon_q$. We find that the coefficient of the lightest state is approximately an order of magnitude higher than the coefficient of the heaviest state, the suppression mainly due to the difference in masses between both states. Thus the sum of Eq.~(\ref{eq-c3333}) is dominated by the lightest state. As expected, larger $\epsilon_q$ leads to larger $C_U^{(n)}$, whereas the dispersion of points is generated by the random variation of the other parameters of the model. A similar dependence is found for $\epsilon_\ell$.

\begin{figure}[h]
\includegraphics[width=0.8\textwidth ]{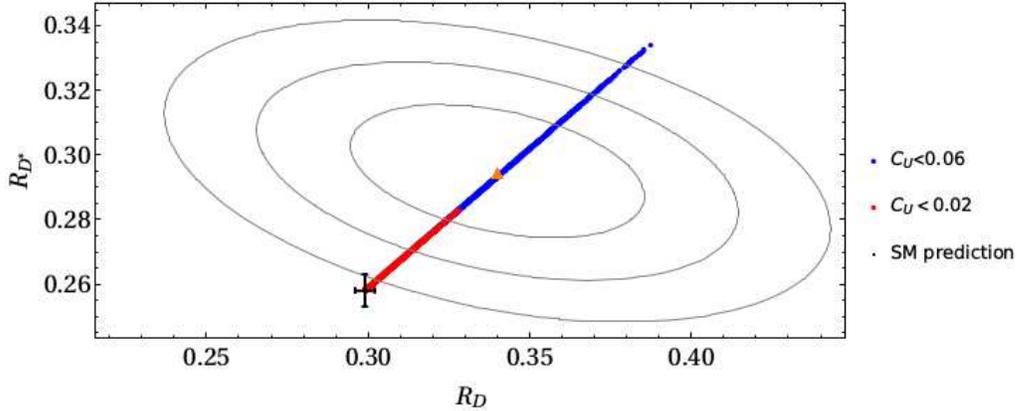}
\centering
\caption{Theoretical predictions for $R_D$ and $R_{D^*}$ for our model. In red are plotted those points that agree with $\sum_nC_U^{(n)}\leq0.02$, consistent with current bounds on LFU violation in tau decays. In blue, we plot the rest of points, up to $\sum_nC_U^{(n)} = 0.06$. Also plotted is the experimental point along with the confidence ellipses for coverage probabilities of 68.27\%, 95.45\% and 99.73\% (1 to 3 $\sigma$)}
\label{ellipses}
\end{figure}

To obtain the coefficient of $R_{D^{(*)}}$ we sum both contributions, and multiply by a random factor $c$ to account for the factors $c_{32}/c_{33}(1-V^*_{tb}g_{L23}/V^*_{ts}g_{L33})$ on Eq.~(\ref{eq-c3233}), that is: $c(C_{U_1}^{(1)}+C_{U_1}^{(2)})$, with $|c|<3$. In Fig.~\ref{ellipses} we present a plot of $R_{D^*}$ vs $R_{D}$. In it, we show the SM prediction, the world average for experimental values, along with confidence ellipses for 1, 2 and 3 $\sigma$. We present our prediction in two groups, one such that $\sum_nC_U^{(n)} \leq 0.02$, to be consistent with LFU in $\tau$ decays, and the rest of the points up to $\sum_nC_U^{(n)} = 0.06$. As can be seen in Fig.~\ref{cus}, this coefficient can actually reach values of order 0.1, we do not show points with $\sum_nC_U^{(n)}\in[0.06,0.1]$. One can see that many points lie in the $1\sigma$ region without violating the bounds from $g_\tau^W$. On the other hand, the bound on $\xi$ from $g_\nu^Z$ is very stringent, we have checked that only 6\% of the points satisfy this bound, almost reaching the border of the 1 $\sigma$ ellipse from below.

%\subsection{Predictions for $R_{K^{(*)}}$}\label{sec-RKn}

\subsection{Phenomenology of the pNGB scalars}
Let us start with the pNGB Higgs. Since the invariants of table~\ref{t-inv1} are the same as for the MCHM with fermions embedded in the fundamental representation of SO(5), the Higgs phenomenology is similar to that case. We will not describe it here, as it has been extensively discussed in the literature, see for instance Refs.~\cite{Contino:2006qr,Carmona:2013cq,Carena:2014ria}, and references therein.

The presence of the $P$-symmetry has important consequences for the phenomenology of $\bar S_1$ and $\varphi$. Let us consider two different cases: first the situation with no elementary $\nu_R$, and second the case with an elementary $\nu_R$ with even parity (+1). In the first case, it is not possible to write a gauge invariant operator, $P$-even , with SM fields and just one power of either $\bar S_1$ or $\varphi$. The lowest dimensional operators with these fields have dimension six and contain both fields, $\varphi$ and $\bar S_1$, they are: ${\cal O}_{q\ell}=\partial_\mu\varphi \bar q_L\bar S_1^*\gamma^\mu \ell_L$ and ${\cal O}_{qe}=\varphi\bar q_L\bar S_1^*H e_R$. Depending on the relation between $m_\varphi$ and $m_{\bar S_1}$, they can mediate either the decay $\varphi\to \bar S_1^*\bar q\ell$, or the decay $\bar S_1\to\varphi\bar q\ell$. The first case leads to a stable particle with electric charge and color: $\bar S_1$, and is not phenomenologically viable. The second case is much more interesting because $\varphi$ is stable and can be a good dark matter candidate (see also Ref.~\cite{Cline:2017aed} for another scenario solving $R_{K^{(*)}}$ and with a dark matter candidate). In Fig.~\ref{scalmass} we have shown the spectrum of these states, showing that both situations are possible in the model. 

Adding an elementary $\nu_R$ even under $P$ allows to include a dimension four operator: ${\cal O}_{u\nu}=\bar u_R\bar S_1^*\nu_R$ that can mediate $\bar S_1$ decay: $\bar S_1^*\to t\nu$. In this case $\varphi$ decays also, mediated by its interactions with $\bar S_1$: $\varphi\to \bar S_1^*\bar q\ell\to t\nu\bar q\ell$, with $\bar S_1$ off-shell for $m_\varphi<m_{\bar S_1}$. This $\nu_R$ can not have Yukawa interactions with $\ell$, since the operator $\bar\ell_L H\nu_R$ is odd under $P$. 

Let us discuss briefly the creation of $\bar S_1$ and $\varphi$ at LHC. $\bar S_1$, being a color triplet, can be created in pairs by QCD interactions: $gg\to \bar S_1\bar S_1^*$. This is the main creation channel at LHC. As discussed in the previous paragraphs, the final state is model dependent. In the interesting case of a stable $\varphi$, one could get a final state with a pair of quarks and leptons of the third generation as well as two scalar singlets: $q\bar q\ell\bar\ell\varphi\varphi$, with the singlets giving missing energy. $\varphi$ could be created in pairs through a dimension six operator as $G_{\mu\nu}G^{\mu\nu}\varphi^2$, or in association with $\bar S_1$, for example in $qZ^*\to\bar S_1^*\varphi\ell\to q\varphi\varphi\ell\bar\ell$, with the virtual $Z^*$ emitted from the initial proton. A detailed study of these processes is beyond the scope of this work.

Direct searches of $\bar S_1$ at LHC give bounds on $m_{\bar S_1}$ of order $\sim1$~TeV, however these bounds depend on which are the dominant decay channels. Since in the present model the decay channels are model dependent, a dedicated analysis must be done for the different cases.

\subsection{Case without $P$-symmetry}\label{sec-noP}
Another possible scenario is the case where $P$-symmetry is violated by the interactions with the elementary sector. There are several possibilities for this violation, we have studied the case where the elementary fermions $u_R$ and $\ell_L$ interact simultaneously with {\bf10} and {\bf10}', respectively contained in {\bf11} and {\bf55} of {\bf66}, as shown in Eq.~(\ref{eq-embed2}). Since {\bf11} is odd and {\bf55} is even under $P$, it is not possible to assign a well defined parity to the elementary fermions such that the elementary-composite interactions are invariant under $P$. It is also possible to break this symmetry with the other elementary fermions, but since their mixing with the SCFT is much smaller than the mixing of $u_R$ and $\ell_L$ of the third generation, we have not studied them.

In this case there are many new operators, as for example dimension one and dimension three operators: $\varphi$, $\varphi|H|^2$ and $\varphi|\bar S_1|^2$, and the dimension five operators: $\varphi\bar\psi\pslash\psi$ (with $\psi$ being any elementary fermion), $\bar q_L\bar S_1^*\pslash\ell_L$ and $\bar u_R\bar S_1^*H^\dagger\ell_L$. The presence of the operators of dimension one and three change drastically the potential, since a vev for $\varphi$ is generated, that can be estimated to be of order $f$. We have computed the one-loop potential to all orders in $\varphi$ and $H$ in the cases without $P$-symmetry, confirming that, in the absence of tuning $\langle\varphi\rangle\sim f$. This vev has a number of new effects, as: it gives large contributions to the EW scale, it induces mixing between $\varphi$ and $h$, it opens new decay channels for $\varphi$ and $\bar S_1$.

The amount of $P$ violation can be controlled by the mixing of $u_R$ with {\bf10}', as well as the mixing of $\ell_L$ with {\bf10}, that we will call $\hat\lambda_u$ and $\hat\lambda_\ell$ respectively ($\lambda_u$ and $\lambda_\ell$ are for the mixings used in the previous sections of the article: {\bf10} for $u$ and {\bf10}' for $\ell_L$). In the limit $\hat\lambda_u=\hat\lambda_\ell=0$ the symmetry is recovered, while taking these mixings small the violation of $P$ is suppressed, and one can obtain $\langle\varphi\rangle\ll f$. As discussed below Eq.~(\ref{eq-linearint}), under some suitable conditions the size of $\lambda_\psi$ is determined by the anomalous dimension of the corresponding SCFT operator ${\cal O}_\psi^{\rm SCFT}$. At the UV scale where Eq.~(\ref{eq-linearint}) is defined ${\cal O}_\psi^{\rm SCFT}$ transforms linearly with a representation of SO(12), thus $\lambda_u=\hat\lambda_u$ and $\lambda_\ell=\hat\lambda_\ell$ at the UV scale. At low IR scales SO(12) is spontaneously broken, allowing a different evolution of $\lambda_\psi$ and $\hat\lambda_\psi$, however, since in our scenario the EW scale is taken only one order of magnitude smaller than that infrared (IR) scale, the window for running is rather small and it is not natural to expect a large hierarchy between $\lambda_\psi$ and $\hat\lambda_\psi$ at the EW scale. Therefore the scenario with small violation of $P$ requires some extra tuning.

We will not elaborate more on this interesting case, and leave it for future work.

%%%%%%%%%%%%%%%%%%%%%%%%%%%%%%%%%%%%%%%%%%%%%%%%%%%%%%%%%%%%%%%%
\section{Conclusions}\label{sec-conclusions}
We have considered a strongly coupled field theory with a unified global symmetry group SO(12) spontaneously broken to SO(11) by the strong dynamics. The breaking SO(12)$\to$SO(11) has the following properties: it contains the SM gauge symmetry and the custodial symmetry, it develops a set of NGBs that include the Higgs, an $\bar S_1$ leptoquark and a SM singlet $\varphi$, and it contains massive spin one states that can be identified with $U_1$ leptoquarks addressing the $B$-anomalies. We have shown that an anarchic flavor structure of the SCFT, together with partial compositeness from linear mixing, can reproduce the SM spectrum and CKM, simultaneously leading to a suitable flavor pattern of couplings of $U_1$. In particular, we have shown that to reproduce the shift in $R_{D^{(*)}}$ a large degree of compositeness of $\ell_L$ and $q_L$ of the third generation is required. This configuration could induce large corrections to $Z$ couplings of $\tau_L$ and $b_L$, that are in agreement with the SM at the per mil level, however we have shown that it is possible to protect those couplings with a well known $LR$ symmetry, by properly choosing the representations of the fermionic operators under the global symmetry of the SCFT. We have shown that $q_L$, $\ell_L$ and $u_R$ can be embedded in the adjoint representation of SO(12), the {\bf66}, whereas $d_R$ and $e_R$ can be embedded in the representation {\bf220}. The elementary-composite interactions, dominated by the third generation, generate a potential for the NGBs at one-loop level, that can trigger EWSB and give masses to the extra NGBs. To obtain the suppression in the masses of the bottom quark and tau, the mixing of $b_R$ and $\tau_R$ must be small, leading to a suppression in the coupling of $U_1$ with the Right handed currents, and realizing the scenario in which only Left handed currents interact with $U_1$.

We have shown an explicit realization of the SCFT dynamics in terms of a weakly coupled theory of resonances. For that we have built an effective low energy theory containing the lowest level of resonances by making use of a three-site theory. This description allows to compute the spectrum and couplings of the resonances, as well as the one-loop potential that is finite. Choosing a large degree of compositeness for $q_L$, $\ell_L$ and $u_R$ of the third generation, we have scanned the parameter space of the three-site theory within a natural region, obtaining a large set of points with EWSB, as well as the right spectrum of SM states. We obtained masses of $\bar S_1$ and $\varphi$ of order $0.2-2$~TeV, and a lightest $U_1$ with mass of order $1-3$~TeV. We computed the couplings of $U_1$ and showed that it is possible to obtain the proper correction to $R_{D^{(*)}}$, without conflict with other observables as the $W$ coupling to $\tau_L$. The anomalies in $R_{K^{(*)}}$ can be solved by properly choosing a small mixing for $q_L$ and $\ell_L$ of the second generation, estimates of the couplings with $U_1$ are given in Ap.~\ref{ap-coup}. We have also shown that the corrections to $Zb_L\bar b_L$ and $Z\tau_L\bar\tau_L$ are of order $0.1\%$ for $\xi\lesssim 0.1$. On the other hand, since there is no protection for $g_\nu^Z$, bounds from this coupling require $\xi\lesssim 0.02$, increasing the amount of tuning and introducing some tension with $R_{D^{(*)}}$. In fact the points with $\xi\lesssim 0.02$ do not enter into de $1\sigma$ region, such that: an improvement of the precission of $R_{D^{(*)}}$, with the same central value, could not be explained in the present model. For the points of the parameter space that induce the proper shift on $R_{D^{(*)}}$, the corrections to $g^W_\tau$ almost saturate the bounds, thus in the present model one can expect to measure deviations in $g^W_\tau$ if measurements of this coupling increase their precision.

We have discussed very briefly the phenomenology of the new scalar states at LHC, finding a very rich set of signals. Since the phenomenology depends on whether the $P$-symmetry, as well as a light elementary Right-handed neutrino with parity +1, are present or not, a detailed study of the production and detection of $\bar S_1$ and $\varphi$ at LHC could allow to distinguish the different realizations of the model. Besides, the size and flavor structure of the couplings are fixed, giving a rather predictive scenario. Although such study is beyond the scope of this work, a careful analysis of direct signals of new physics that could be related with the $B$-anomalies must be done, particularly at LHC. We find this avenue very interesting and leave its study for future work.

Finally, we stress that the coset SO(11)/SO(10) is big enough to contain the SM and the custodial symmetry, develop $H$ as a NGB and generate a $U_1$ leptoquark. However, since in this case $U_1$ is associated to broken generators, its mass results heavier than the lightest resonances associated to $Z$ and $W$. EWPT give lower bounds on the later of order 2-3~TeV, thus $m_{U_1}\gtrsim 4-6$~TeV. These values of $m_{U_1}$ give an extra suppression to the contribution to $R_{D^{(*)}}$, that results approximately a factor 4 smaller than what is needed to fit the anomalies. It could be interesting to study the possibility of finding a group smaller than SO(12) that could do the job.

\section*{Acknowledgements}
We thank Eduardo Andr\'es for help with group theory, Bartosz Fornal for discussions, Mariel Estevez for help with analysis of $Z'$-states, and CONICET Argentina for financial support with PIP-0299.

%%%%%%%%%%%%%%%%%%%%%%%%%%%%%%%%%%%%%%%%%%%%%%%%%%%%%%%%%%%%%%%%

\appendix
\section{Representations of SO(12)}\label{ap-SO(12)}
In this appendix we give a brief description of the algebra, as well as the lowest dimensional representations, of the group SO(12).
A simple basis for the algebra of SO(12) in the fundamental representation is given by the set of generators $\{T_{\ell m},\ell<m=2,\dots 12\}$, with coefficients:
\begin{equation}
  ({\cal T}_{\ell, m})_{jk}=i(\delta_{\ell j}\delta_{mk}-\delta_{mj}\delta_{\ell k}) \ , \qquad l<m \ .
  \label{tij}
\end{equation}

An SO(11) subgroup can be defined by choosing a vector $\hat{n}$ to point in a direction of the 12-dimensional space. For instance, selecting the twelfth coordinate, $\hat{n} = \hat{e}_{12}$, the algebra of SO(11) is defined by generators as in Eq.~(\ref{tij}) with indices different from ``12''. Inside SO(11), we can define the subgroup SO(4)$\times$SO(6), where we will embed SU(2)$_L \times$SU(2)$_R\times$SU(3)$_C\times$U(1)$_X$. The SO(4) algebra is defined by allowing indices to run from 1 to 4, while the SO(6) algebra by those indices between 6 and 11.  

The algebra of SU(2)$_L\times$SU(2)$_R$ inside SO(4) can be defined by:
\begin{align}
&T^L_1=-\frac{1}{2}({\cal T}_{1,4}+{\cal T}_{2,3}) \ ,&
&T^L_2=\frac{1}{2}({\cal T}_{1,3}-{\cal T}_{2,4}) \ , &
&T^L_3=-\frac{1}{2}({\cal T}_{1,2}+{\cal T}_{3,4}) \ , \nonumber
\\
&T^R_1=\frac{1}{2}({\cal T}_{1,4}-{\cal T}_{2,3}) \ ,&
&T^R_2=\frac{1}{2}({\cal T}_{1,3}+{\cal T}_{2,4}) \ , &
&T^R_3=-\frac{1}{2}({\cal T}_{1,2}-{\cal T}_{3,4}) \ , \nonumber
\label{alg-su23}
\end{align}
An algebra of SU(3)$\times$U(1) inside SO(6) can be defined by:
\begin{align}
&T^{\rm SU(3)}_1=\frac{1}{2}({\cal T}_{8,11}-{\cal T}_{9,10}) \ ,&
&T^{\rm SU(3)}_2=\frac{1}{2}({\cal T}_{8,10}+{\cal T}_{9,11}) \ ,\nonumber \\
&T^{\rm SU(3)}_3=\frac{1}{2}(-{\cal T}_{8,9}+{\cal T}_{10,11}) \ ,&
&T^{\rm SU(3)}_4=\frac{1}{2}({\cal T}_{6,11}-{\cal T}_{7,10}) \ ,\nonumber \\
&T^{\rm SU(3)}_5=\frac{1}{2}({\cal T}_{6,10}+{\cal T}_{7,11}) \ ,&
&T^{\rm SU(3)}_6=\frac{1}{2}({\cal T}_{6,9}-{\cal T}_{7,8}) \ ,\nonumber \\
&T^{\rm SU(3)}_7=\frac{1}{2}({\cal T}_{6,8}+{\cal T}_{7,9}) \ ,&
&T^{\rm SU(3)}_8=\frac{1}{2\sqrt{3}}(-2{\cal T}_{6,7}+{\cal T}_{8,9}+{\cal T}_{10,11}) \ ,\nonumber\\
&T^{\rm U(1)}=-4({\cal T}_{6,7}+{\cal T}_{8,9}+{\cal T}_{10,11}) \ , &&
%\label{alg-su31}
\end{align}

We construct the adjoint representation ({\bf 66}) by using the structure constants, or, by using the generators of the algebra as a basis of this vector space. 

The smallest representations of SO(12), and their decompositions under SO(10):
\begin{align}
& {\bf 12}\sim  \mathbf{1} \oplus \mathbf{11}  ,  \nonumber
\\
& {\bf 66}\sim \mathbf{11} \oplus \mathbf{55}  , \nonumber
\label{eq-frep13-1}
\end{align}

%  & {\bf 1} \sim ({\bf 1},{\bf 1},{\bf 1})_{0} \nonumber \\
Decomposing them further under H$_\textrm{min}$ to identify which representations contain SM fermions, we get:
\begin{align}
   {\bf 1} &\sim ({\bf 1},{\bf 1},{\bf 1})_{0} \nonumber \\
   {\bf 11} &\sim ({\bf 3},{\bf 1},{\bf 1})_{1/\sqrt{6}}\oplus({\bf 1},{\bf 2},{\bf 2})_0\oplus({\bf 1},{\bf 1},{\bf 1})_0\oplus{\rm c.c.}\ , \nonumber
\\
 {\bf 55}&\sim ({\bf 3},{\bf 2},{\bf 2})_{1/\sqrt{6}}\oplus({\bf 3},{\bf 1},{\bf 1})_{1/\sqrt{6}}\oplus({\bf 3},{\bf 1},{\bf 1})_{-2/\sqrt{6}}\oplus ({\bf 8},{\bf 1},{\bf 1})_{0}\oplus({\bf 1},{\bf 2},{\bf 2})_0 \\ & \quad\quad\oplus({\bf 1},{\bf 3},{\bf 1})_0\oplus(\mathbf{1},\mathbf{1},\mathbf{3})\oplus({\bf 1},{\bf 1},{\bf 1})_0\oplus{\rm c.c.}\ , \nonumber
%& 286\sim ({\bf 1},{\bf 1},{\bf 1})_6\oplus({\bf 1},{\bf 3},{\bf 1})_0\oplus({\bf 8},{\bf 3},{\bf 1})_0\oplus({\bf 8},{\bf 2},{\bf 2})_0\oplus({\bf 1},{\bf 2},{\bf 2})_0 \nonumber 
%\\
%&\qquad \oplus({\bf 3},{\bf 1},{\bf 1})_2\oplus({\bf 6},{\bf 1},{\bf 1})_{-2}\oplus({\bf 3},{\bf 3},{\bf 1})_{-4}\oplus({\bf 3},{\bf 2},{\bf 2})_{-4}\oplus({\bf 3},{\bf 3},{\bf 1})_2\oplus({\bf 3},{\bf 1},{\bf 3})_2\oplus({\bf 3},{\bf 2},{\bf 2})_2\oplus({\bf 3},{\bf 4},{\bf 2})_2 \oplus{\rm c.c.} \ ,
%\\
%& 286\sim ({\bf 3},{\bf 1},{\bf 1})_2\oplus({\bf 6},{\bf 1},{\bf 1})_{-2}\oplus({\bf 3},{\bf 3},{\bf 1})_{-4}\oplus({\bf 3},{\bf 2},{\bf 2})_{-4}\oplus({\bf 3},{\bf 3},{\bf 1})_2\oplus({\bf 3},{\bf 1},{\bf 3})_2\oplus({\bf 3},{\bf 2},{\bf 2})_2\oplus({\bf 3},{\bf 4},{\bf 2})_2 \oplus{\rm c.c.} \nonumber 
%\\
%&\qquad\oplus({\bf 8},{\bf 3},{\bf 1})_0\oplus({\bf 8},{\bf 2},{\bf 2})_0 \oplus({\bf 1},{\bf 1},{\bf 1})_6\oplus({\bf 1},{\bf 3},{\bf 1})_0\oplus({\bf 1},{\bf 2},{\bf 2})_0 \ ,
%\\
%\label{eq-frep13-2}
\end{align}
the complex conjugate representations must be added only when they are not equivalent to the original one.
As we need to consider a parity transformation in order to forbid odd terms in the pNGB potential, we will have to extend the group from SO(12) to O(12), as this transformation has a determinant equal to -1. We wish to make the pNGB states odd under this parity. One way to achieve this is to make this parity act over the fundamental representation ${\bf 12}$ as:
\begin{align}
  & P\, {\bf 11} = {\bf 11} \\ 
  & P\, {\bf 1} = -{\bf 1}
\end{align}
The way to represent this parity transformation in the basis here defined would be as a diagonal matrix with its first 11 entries +1, and the last entry -1. As the adjoint representation ${\bf 66}$ can be built with the product of two ${\bf 12}$ representations, we can find how this parity acts on the adjoint by decomposing the product of representations: 
\begin{align}
 &  P\, {\bf 55} = {\bf 55} \\ 
  & P\, {\bf 11} = -{\bf 11}
  \end{align}
And the pNGB are inside this ${\bf 11}$ representation within the adjoint, so they will be odd under this parity.

\section{Mass matrices}\label{ap-mm}
In this appendix we show the mass matrices in the three-site model, for the fermions we consider just one generation and we do not include $d_R$, nor $e_R$ partners, since their mixings are small. 

For the up type quarks we get a nine by nine matrix, as there are four elements inside the adjoint representation with the same SM quantum numbers as this fermion. In the basis where we first put the elementary fermion, then the four representations of site 1, and then those of site 2, we get:
\begin{equation}
  M_u =  \begin{pmatrix} 0 & 0 & -i\lambda_u s_v/\sqrt{2} & i \lambda_u s_v/\sqrt{2} & \lambda_u c_v & 0 & 0 & 0 &  0 \\
  0 & m_1 & 0 & 0 & 0 & \lambda_{1,2} & 0 & 0 &  0 \\
  \lambda_q s_{v/2}^2 & 0 & m_1 & 0 & 0 & 0 & \lambda_{1,2} & 0 &  0 \\
  \lambda_q c_{v/2}^2 & 0 & 0 & m_1 & 0 & 0 & 0 & \lambda_{1,2} &  0 \\
  -i \lambda_q s_v /\sqrt{2} & 0 & 0 & 0 & m_1 & 0 & 0 & 0 &  \lambda_{1,2} \\
  0 & \lambda_{1,2} & 0 & 0 & 0 & m_{2,\bf55} & 0 & 0 &  0 \\
  0 & 0 & \lambda_{1,2} & 0 & 0 & 0 & m_{2,\bf55} & 0 &  0 \\
  0 & 0 & 0 & \lambda_{1,2} & 0 & 0 & 0 & m_{2,\bf55} &  0 \\
  0 & 0 & 0 & 0 & \lambda_{1,2} & 0 & 0 & 0 &  m_{2,\bf11} 
  \end{pmatrix}
\end{equation}

For the down type quarks, the matrix is a smaller three by three matrix as there is only one representation inside ${\bf66}$ that contains the adequate quantum numbers. In the basis where we order the three elements as site 0, 1 and 2 respectively, we get:
\begin{equation}
  M_d = \begin{pmatrix}
    0 & 0 & 0 \\
    \lambda_q & m_1 & \lambda_{1,2} \\
    0 & \lambda_{1,2} & m_{2,\bf55} 
 \end{pmatrix}
\end{equation}
There is a row of zeros because there are no mixing for $d_R$. 

For the charged lepton we also get a nine-by-nine mass matrix:
\begin{equation}
  M_e =  \begin{pmatrix} 0 & 0 & 0 & 0 & 0 & 0 & 0 & 0 &  0 \\
  - \lambda_\ell  & m_1 & 0 & 0 & 0 & \lambda_{1,2} & 0 & 0 &  0 \\
  0 & 0 & m_1 & 0 & 0 & 0 & \lambda_{1,2} & 0 &  0 \\
  0 & 0 & 0 & m_1 & 0 & 0 & 0 & \lambda_{1,2} &  0 \\
  0 & 0 & 0 & 0 & m_1 & 0 & 0 & 0 &  \lambda_{1,2} \\
  0 & \lambda_{1,2} & 0 & 0 & 0 & m_{2,\bf55} & 0 & 0 &  0 \\
  0 & 0 & \lambda_{1,2} & 0 & 0 & 0 & m_{2,\bf55} & 0 &  0 \\
  0 & 0 & 0 & \lambda_{1,2} & 0 & 0 & 0 & m_{2,\bf55} &  0 \\
  0 & 0 & 0 & 0 & \lambda_{1,2} & 0 & 0 & 0 &  m_{2,\bf11} 
  \end{pmatrix}
\end{equation}

We do the same for the bosonic resonances. For the $U_1$ state, we get a three by three matrix, as there is, in site 1 both broken and unbroken generators identified with it, but in the site 2 only an unbroken one. We order the basis as the two unbroken generators in site 1 and 2, followed by the broken one in site 1. We get
\begin{equation}
  M_{U_1} = \frac12 \begin{pmatrix}
   g1^2 \left( f_1^2 + f_2^2\right)  & - f_2^2 g_1 g_2 & 0 \\
   - f_2^2 g_1 g_2 & f_2^2 g_2^2 & 0  \\
 0 & 0 & g_1^2 \left(f_1^2 + f_2^2\right)  
    \end{pmatrix}
\end{equation}

For $Z$ resonances we get a 14-by-14 matrix, as there are 8 elements in the algebra associated with the Z quantum numbers, but three of those belong to the unbroken generators, which are not present in site 2. Adding a source in the site 0, we get 14 degrees of freedom. We do not show the matrix because it is too large.

\section{Numerical estimates of the Right-handed $U_1$ couplings}\label{ap-coup}
The couplings of the $U_1$ leptoquarks to elementary fermions were estimated in Eq.~(\ref{eq-gU1}). Making use of Eqs.~(\ref{eq-mixq}) and (\ref{eq-mixl}), and assuming that all the couplings between the resonances are of the same order (this is known as the {\it one coupling} scenario), we obtain for the Right-handed couplings:
\begin{equation}\label{eq-estgR}
g_R^{(n)} \sim \epsilon_{u3}
\begin{pmatrix}
\frac{m_dm_e}{\lambda_C^3v^2}\frac{1}{\epsilon_{\ell 1}}   
&
\frac{m_dm_\mu}{\lambda_C^3v^2}\frac{1}{\epsilon_{\ell 2}}   
&
\frac{m_dm_\tau}{\lambda_C^3v^2}\frac{1}{\epsilon_{\ell 3}}   
\\
\frac{m_sm_e}{\lambda_C^2v^2}\frac{1}{\epsilon_{\ell 1}}   
&
\frac{m_sm_\mu}{\lambda_C^2v^2}\frac{1}{\epsilon_{\ell 2}}   
&
\frac{m_sm_\tau}{\lambda_C^2v^2}\frac{1}{\epsilon_{\ell 3}}   
\\
\frac{m_bm_e}{v^2}\frac{1}{\epsilon_{\ell 1}}   
&
\frac{m_bm_\mu}{v^2}\frac{1}{\epsilon_{\ell 2}}   
&
\frac{m_bm_\tau}{v^2}\frac{1}{\epsilon_{\ell 3}}   
\end{pmatrix}
\end{equation}

The experimental values of $R_{D^{(*)}}$ and $R_{K^{(*)}}$ can be reproduced by taking $\epsilon_{\ell 3}\sim \epsilon_{u3}m_{U_1}/$~TeV and $\epsilon_{\ell 2}/\epsilon_{\ell 3}\sim 0.2$. %As discussed at the end of sec.~\ref{sec-Ba}, Eq.~(\ref{eq-mixl}) leads to $\epsilon_{e3}\simeq 0.7\times10^{-2}/g_1*$ and $\epsilon_{e2}\simeq 0.4\times10^{-3}/g_*$.
As discussed at the end of sec.~\ref{sec-Ba}, $\epsilon_{\ell 1}$ is not fixed by the $B$-anomalies, as long as $\epsilon_{\ell 1}\ll \epsilon_{\ell 2}$. Taking for simplicity $\epsilon_{\ell 1}\sim \epsilon_{e 1}\sim (m_e/vg_*)^{1/2}$ and $\epsilon_{u3}\sim\epsilon_{q3}\sim 1/\sqrt{g_*}$, Eq.~(\ref{eq-estgR}) takes the values:
\begin{equation}
g_R^{(n)} \sim \left(
\begin{array}{ccc}
10^{-6} & 2\times 10^{-6}x & 5\times 10^{-6}x \\
6\times 10^{-6} & 9\times 10^{-6}x & 3\times 10^{-5}x \\
10^{-5} & 2\times 10^{-5}x & 6\times 10^{-5}x 
\end{array}
\right)
\ ,\qquad
x=\frac{\rm TeV}{m_{U_1}} \ .
\end{equation}
The Right-handed couplings are much smaller than the Left-handed ones, thus with good accuracy one can neglect $g_R$ and consider just the interactions with the Left-handed currents.

\section{A model in 5 dimensions}\label{ap-extradim}
An holographic dual of the SCFT described in sec.~\ref{sec-3sites} can be obtained by working in a theory with extra dimensions. We consider a 5D spacetime, with a compact extra dimension and an AdS$_5$ geometry. The extra dimension has two boundaries, respectively called UV-boundary and IR-boundary, leading to the well known Randall-Sundrum model. The IR scale is taken of order TeV, whereas the UV one is of the order of the Planck scale.

There is an SO(12) gauge symmetry in the bulk, broken to SO(11) in the IR by boundary conditions, and to the SM gauge symmetry group in the UV also by boundary conditions. A zero mode of the fifth component of the gauge field, $A_5$, survives and leads to the NGBs, that transform in the fundamental representation of SO(11). This is very similar to the MCHM arising from SO(5)/SO(4).

There is a 5D fermion for each SM fermion multiplet of the SM. We will describe only those associated to $q$, $u$ and $\ell$ of the third generation, that are embedded in the representation {\bf66} of SO(12). We will use large letters for the 5D fields: $Q$, $U$ and $L$. To simplify the explanation of the boundary conditions of the 5D fermions, we find it useful to define the following notation for the decomposition of representations of SO(11) under G$_{\rm SM}$:
\begin{align}
&{\bf 55}\sim {\bf r}_q \oplus \tilde {\bf r}_q \ , \qquad {\bf r}_q = ({\bf 3},{\bf 2})_{1/6}\oplus{\rm c.c.} \ , 
\nonumber \\
&{\bf 55}\sim {\bf r}_\ell \oplus \tilde {\bf r}_\ell \ , \qquad {\bf r}_\ell = ({\bf 1},{\bf 2})_{-1/2} \ , 
\nonumber \\
&{\bf 11}\sim {\bf r}_u \oplus \tilde {\bf r}_u \ , \qquad {\bf r}_u = ({\bf 3},{\bf 1})_{2/3}\oplus{\rm c.c.} \ . 
\end{align}
$\tilde{\bf r}_q$, $\tilde{\bf r}_\ell$ and $\tilde{\bf r}_u$ are reducible representations of G$_{\rm SM}$, that can be expressed straightforwardly in terms of irreducible ones, see for example Eq.~(\ref{eq-rep55}) for the decomposition of {\bf55} under H$_{\rm min}$.

The boundary conditions of the 5D fermions can be taken as:
\begin{align}
&Q_L=\left[\begin{array}{l}
Q_L^{\bf 55}=\left[\begin{array}{l} Q_L^{{\bf r}_q}(++)\\Q_L^{\tilde{\bf r}_q}(-+)\end{array}\right]
\\
Q_L^{\bf 11}(--)
\end{array}\right] \ , 
\qquad
Q_R=\left[\begin{array}{l}
Q_R^{\bf 55}=\left[\begin{array}{l} Q_R^{{\bf r}_q}(--)\\Q_R^{\tilde{\bf r}_q}(+-)\end{array}\right]
\\
Q_R^{\bf 11}(++)
\end{array}\right] \ ,
\nonumber \\
&L_L=\left[\begin{array}{l}
L_L^{\bf 55}=\left[\begin{array}{l} L_L^{{\bf r}_\ell}(+-)\\L_L^{\tilde{\bf r}_\ell}(--)\end{array}\right]
\\
L_L^{\bf 11}(-+)
\end{array}\right] \ , 
\qquad
L_R=\left[\begin{array}{l}
L_R^{\bf 55}=\left[\begin{array}{l} L_R^{{\bf r}_\ell}(-+)\\L_R^{\tilde{\bf r}_\ell}(++)\end{array}\right]
\\
L_R^{\bf 11}(+-)
\end{array}\right]
\ ,
\nonumber \\
&U_L=\left[\begin{array}{l}
U_L^{\bf 55}(+-)
\\
U_L^{\bf 11}=\left[\begin{array}{l} U_L^{{\bf r}_u}(-+)\\U_L^{\tilde{\bf r}_u}(++)\end{array}\right]
\end{array}\right] \ , 
\qquad
U_R=\left[\begin{array}{l}
U_R^{\bf 55}(-+)
\\
U_R^{\bf 11}=\left[\begin{array}{l} L_R^{{\bf r}_u}(+-)\\L_R^{\tilde{\bf r}_u}(--)\end{array}\right]
\end{array}\right] \ .
\end{align}
Notice that, since SO(12) is broken to SO(11) in the IR boundary, the boundary conditions of different SO(11) components of the 5D fermions can be different in the IR. The boundary conditions on the UV can be understood easily by making use of the holographic approach~\cite{Contino:2004vy}: describing $Q$ and $L$ in terms of Left-handed sources localized on the UV, and $U$ in terms of Right-handed ones, only a set of UV-sources that can be matched with the SM ones are dynamical, {\it i.e.}: they have (+) UV boundary conditions. 

We include also a 4D chiral fermion localized in the IR boundary: $\chi_L$, transforming as a {\bf55}, as well as the following mass terms localized on the IR:
\begin{equation}\label{eq-L5DIR}
{\cal L}_{\rm IR}= m_u\bar Q_L^{\bf55} U_R^{\bf55}+\tilde m_u\bar \chi_L U_R^{\bf55}+m_\ell\bar \chi_L L_R^{\bf55}+m_1\bar Q_L^{\bf55} L_R^{\bf55}+m_2\bar L_L^{\bf11} Q_R^{\bf11}+{\rm h.c.} \ ,
\end{equation}
that are compatible with the SO(11) symmetry of the IR boundary.

The fermions with (++) boundary conditions lead to chiral 0-modes. After taking into account the presence of $\chi_L$ and Eq.~(\ref{eq-L5DIR}), there remain three massless chiral fields that can be identified with $q_L$, $u_R$ and $\ell_L$ of the SM.

As usual in this kind of theories, the degree of compositeness of the 0-modes is controlled by the bulk mass of the 5D fermions. The form factors can be calculated by integration of the bulk degrees of freedom, and they can be expressed in terms of Bessel functions~\cite{Contino:2004vy,Agashe:2004rs}.

The lightest $U_1$ leptoquarks are given by the lightest Kaluza-Klein states of the 5D gauge field associated to the generators transforming as $({\bf3},{\bf1})_{2/3}$ inside {\bf55}. 

We will not elaborate more on this description.

\bibliographystyle{JHEP}
\bibliography{biblio_v2_2}
\end{document}